\documentclass[twocolumn]{aastex631}
\usepackage{xspace}
\usepackage{placeins}
\usepackage{amsmath}
\newcommand{\swift}{{\it Swift}\xspace}
\newcommand{\nicer}{\textit{NICER}\xspace}
\newcommand{\xmm}{{\it XMM-Newton}\xspace}

\newcommand{\target}{{AT2020ksf}\xspace}

\begin{document}
\title{Delayed X-ray brightening accompanied by variable ionized absorption following a tidal disruption event}
\correspondingauthor{Thomas Wevers}
\email{twevers@stsci.edu}
\author[0000-0002-4043-9400]{T. Wevers}
\affiliation{Space Telescope Science Institute, 3700 San Martin Drive, Baltimore, MD 21218, USA}
\affiliation{European Southern Observatory, Alonso de Córdova 3107, Vitacura, Santiago, Chile}
\author[0000-0002-5063-0751]{M. Guolo}
\affiliation{Department of Physics and Astronomy, Johns Hopkins University, 3400 N. Charles St., Baltimore MD 21218, USA}
\author[0000-0003-1386-7861]{D.R. Pasham}
\affiliation{Kavli Institute for Astrophysics and Space Research, Massachusetts Institute of Technology, Cambridge, MA, USA}
\author[0000-0003-3765-6401]{E.R. Coughlin}
\affiliation{Department of Physics, Syracuse University, Syracuse, NY 13210, USA}
\author[0000-0002-6562-8654]{F. Tombesi}
\affiliation{Physics Department, Tor Vergata University of Rome, Via della Ricerca Scientifica 1, 00133 Rome, Italy}
\affiliation{INAF – Astronomical Observatory of Rome, Via Frascati 33, 00040 Monte Porzio Catone, Italy}
\affiliation{INFN - Rome Tor Vergata, Via della Ricerca Scientifica 1, 00133 Rome, Italy }
\affiliation{Department of Astronomy, University of Maryland, College Park, MD 20742, USA}
\affiliation{NASA Goddard Space Flight Center, Code 662, Greenbelt, MD 20771, USA}

\author[0000-0001-6747-8509]{Y. Yao}
\affiliation{Cahill Center for Astronomy and Astrophysics, California Institute of Technology, Pasadena, CA 91125, USA}
\author[0000-0003-3703-5154]{S. Gezari}
\affiliation{Space Telescope Science Institute, 3700 San Martin Drive, Baltimore, MD 21218, USA}\affiliation{Department of Physics and Astronomy, Johns Hopkins University, 3400 N. Charles St., Baltimore MD 21218, USA}

\begin{abstract}
Supermassive black holes can experience super-Eddington peak mass fallback rates following the tidal disruption of a star. The theoretical expectation is that part of the infalling material is expelled by means of an accretion disk wind, whose observational signature includes blueshifted absorption lines of highly ionized species in X-ray spectra. To date, however, only one such ultra-fast outflow (UFO) has been reported in the tidal disruption event (TDE) ASASSN--14li. Here we report on the discovery of transient absorption-like signature in X-ray spectra of the TDE AT2020ksf/Gaia20cjk (at a redshift of $z$=0.092), following an X-ray brightening $\sim 230$ days after UV/optical peak. 
We find that while no statistically significant absorption features are present initially, they appear on a timescale of several days, and remain detected up to 770 days after peak.
Simple thermal continuum models, combined with a power-law or neutral absorber, do not describe these features well. Adding a partial covering, low velocity ionized absorber improves the fit at early times, but fails at late times. A high velocity (v$_w$ $\sim$ 42\,000 km s$^{-1}$), ionized absorber (ultra-fast outflow) provides a good fit to all data. The few day timescale of variability is consistent with expectations for a clumpy wind. We discuss several scenarios that could explain the X-ray delay, as well as the potential for larger scale wind feedback.
The serendipitous nature of the discovery could suggest a high incidence of UFOs in TDEs, alleviating some of the tension with theoretical expectations. 
\end{abstract}

\section{Introduction} \label{sec:intro}
When a star passes close enough to a massive black hole ($\sim$10$^{4-8}$ M$_{\odot}$) such that the tidal shear across the length of the star exceeds its self-gravity, it will be disrupted \citep{Rees1988, Kochanek94}. Such stellar tidal disruption events (TDEs) have been heralded as ideal systems to study the formation of accretion disks and outflows launched from black holes. Ultra-fast outflows (UFOs, those moving with line of sight velocities $\gtrsim$10\,000 km s$^{-1}$) are especially interesting because they carry mass and momentum away from the hole, aid in the formation of disks and potentially the X-ray corona. Remarkably, they can play a significant role in galaxy feedback by pumping copious amounts of energy into their environment (\citealt{Pounds03, Cappi06, Tombesi10}; see \citealt{King2015} for a review). 

However, to-date only one TDE, ASASSN-14li has been reported to exhibit a UFO signature \citep{Kara18}, although tentative evidence has been found in some other sources (e.g. \citealt{Saxton2012, Lin2015}). Taken at face value, this is surprising within the context of state-of-the-art 3-dimensional general relativistic radiation magnetohydrodynamics (GRRMHD) simulations, which predict that outflows should be ubiquitous in TDEs \citep{Dai2018, Thomsen22}. These UFOs are detectable at UV and/or X-ray wavelengths if the wind favorably intersects with the observer line of sight (e.g. \citealt{Parkinson22}). A mitigating factor to this tension between predictions and observations is the fact that the detection of a UFO signature in X-ray spectra requires high signal-to-noise spectra, which generally implies long exposure times \citep{Kara18}. Such observations are not available for many TDEs \citep{vv21}, hence firm conclusions cannot yet be drawn from the lack of detections. An alternative probe of such disk winds can be found at radio wavelengths, if and when the wind interacts with the circum-nuclear medium, driving a forward shock that triggers synchrotron (radio) emission (e.g. \citealt{Alexander17}). However, this provides a more indirect probe of the dynamical evolution of the bulk wind properties, such as its kinetic power, and it is not possible to derive physical properties such as the density and ionization state of the wind. In recent years, the Neutron Star Interior Composition ExploreR (NICER; \citealt{keith}) has followed up several TDEs with deep and high cadence exposures, yielding high-quality X-ray spectra with $\gtrsim$ a few$\times$(1000-10,000) X-ray photons. Such observations, when carried out systematically, should enable robust constraints on the presence (or absence) of powerful disk winds in TDEs. 

Here we report on the discovery of transient absorption signatures in the X-ray spectra of the TDE AT2020ksf at 230 days and 700 days after peak. Combinations of simple continuum models, such as the combination of a thermal and a power law model or two thermal models, lead to systematic trends in the fit residuals and hence cannot describe the data well. A partial covering, low velocity ionized absorber ({\it warm} absorber) model improves the description of the data at early times, but does not work for the late-time data. Instead, if we model the early and late time spectra with an ultra-fast outflow we can fit both epochs well. The manuscript is organized into four sections. We present the observations and their analysis in Section \ref{sec:results}, discuss the properties of the UFO and their implications in Section \ref{sec:discussion}, and summarize in Section \ref{sec:summary}. In the supplementary materials (Section \ref{sec:data}) we describe the observations, data reduction and additional analysis.

\section{Observations, analysis and results}
\label{sec:results}
Gaia20cjk/AT2020ksf was discovered as a 2 magnitude outburst in the center of a galaxy at redshift $z$=0.092 (luminosity distance of 426 Mpc) on 2020 April 21 (MJD 58960, which is taken as the reference point for all phases and lightcurves) by the Gaia photometric science alerts \citep{Hodgkin21}. An X-ray detection by eROSITA was reported on 2020 November 9 \citep{2020ATel14246....1G}, and a radio detection was reported on 2020 December 21 (MJD 59204, 244 days after discovery) with the Karl G. Jansky Very Large Array (VLA) at 6 GHz, with a flux level of 47$\pm$10 $\mu$Jy, corresponding to a monochromatic luminosity of L$_{\rm radio} = 6 \pm 1 \times 10^{37}$ erg s$^{-1}$ \citep{2021TNSAN..24....1A}. The transient was classified as a tidal disruption event based on optical spectroscopy, showing broad H$\alpha$ and He\,\textsc{ii} $\lambda$4686 emission lines \citep{2020ATel14246....1G}. 

The data presented in this work were acquired by five different instruments: \nicer/XTI (\citealt{keith}; see also \citealt{Mummery23} for a continuum analysis of this work), \xmm/EPIC \citep{xmm}, \swift's X-Ray Telescope (XRT; \citealt{Burrows2005}) and the UV Optical Telescope (UVOT; \citealt{Roming2005}), and {\it Spectrum-Roentgen-Gamma}/eROSITA \citep{Predehl21}. We also use publicly available data from Gaia \citep{Hodgkin21}, the Zwicky Transient Facility (ZTF; \citealt{ztf}) and the Asteroid Terrestrial-impact Last Alert System (ATLAS; \citealt{atlas}) to derive the optical evolution of AT2020ksf, and a Keck Échelle Spectrograph and Imager (ESI) optical spectrum to characterize the host galaxy. The data reduction is described in the Supplementary material (Sec. \ref{sec:data}).

\begin{figure*}
    \centering
    \includegraphics[width=0.95\textwidth]{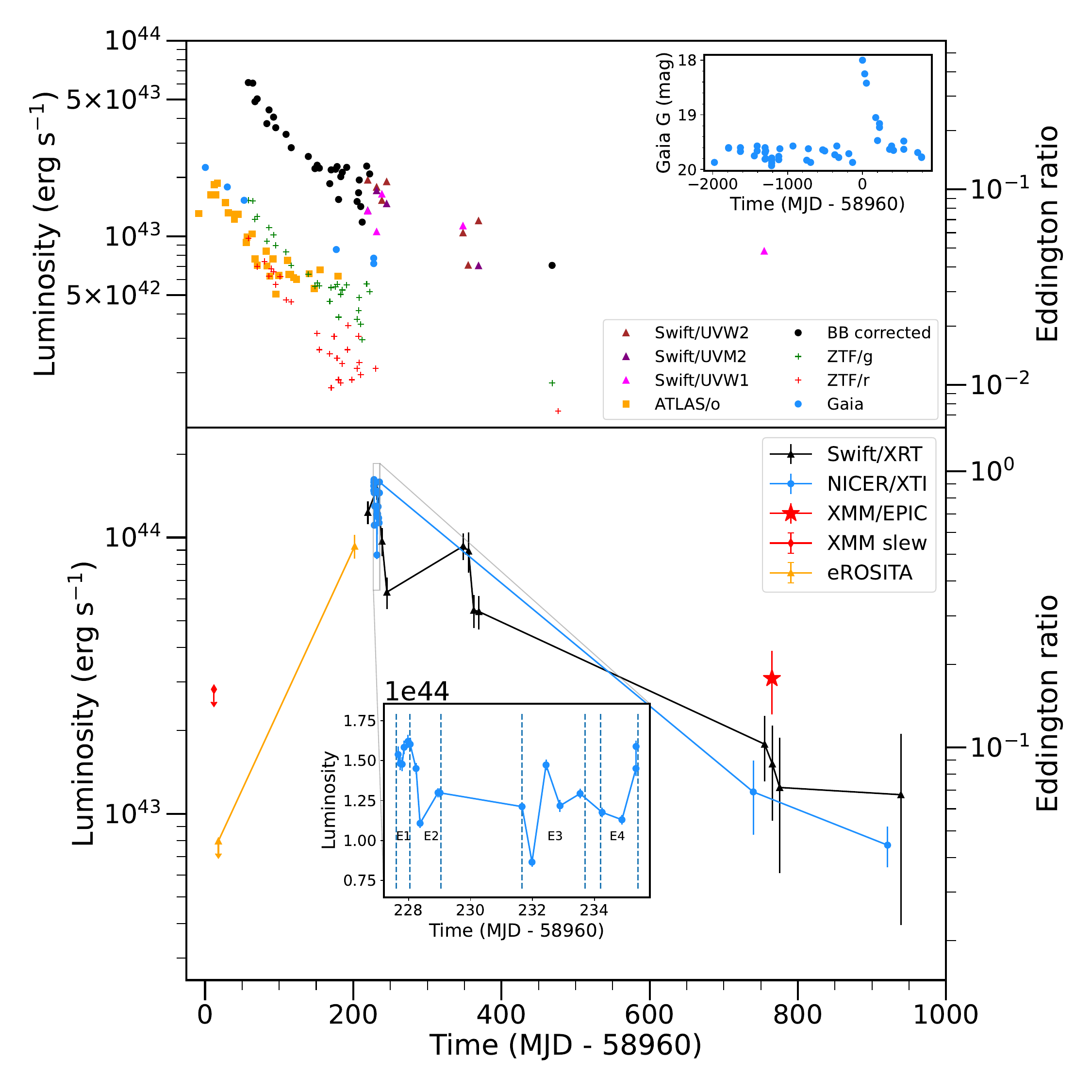}
    \caption{{\bf X-ray and UV lightcurves of AT2020ksf.} The top panel shows the UV/optical evolution in the Swift/UV and various optical bands. The inset shows the Gaia lightcurve up to 2000 days before discovery, indicating no significant host galaxy variability. The black circles show the blackbody-corrected 0.03--3 $\mu$m lightcurve based on the ZTF g-band, assuming a temperature of 20\,000 K. The bottom panel shows the Swift, NICER, XMM-Newton and eROSITA X-ray lightcurves, all computed in the spectral range 0.3--1.1 keV. Peak light in the X-rays is delayed with respect to the UV/optical, as constrained by eROSITA and XMM-Newton slew survey observations (orange triangles and red diamond). Note that when integrating the X-ray spectrum from 1--1000 Ryd, the luminosity remains near the Eddington limit for at least 770 days. The inset shows the NICER high cadence data near the X-ray lightcurve peak. Variability by a factor of 2--3 on several day timescales is present throughout. The NICER epochs that are used for spectral modeling are marked by dashed and dotted lines; for example, E34 signifies that data from E3 and E4 were stacked. }
    \label{fig:lc}
\end{figure*}
\subsection{Light curves and black hole mass}
The UV/optical and X-ray lightcurves are presented in Figure \ref{fig:lc}, in the top and bottom panels respectively. We compute a bolometric correction from the ZTF $g$-band to the 0.03--3 $\mu$m wavelength range (black dots), corresponding to a typical blackbody temperature of 20\,000 K (the correction factor is $\approx$4). With this estimate, the integrated UV/optical emission peaks around 7$\times$10$^{43}$ erg s$^{-1}$ and hence approaches the Eddington limit ($\sim$10$^{44}$ erg s$^{-1}$) at peak for a black hole mass of 10$^{6}$ M$_{\odot}$ (see below). At late times, it flattens out at 5$\times$10$^{42}$ erg s$^{-1}$.

All X-ray lightcurves were converted to the same 0.3--1.1 keV band,  assuming the best-fit spectral models determined in Section \ref{sec:specmod} to translate count rates into luminosities. For the Swift data and NICER E12 spectrum, we adopt the thermal continuum spectral models of Table \ref{tab:continuumfits}. For the E34 and XMM data, a UFO model significantly improves the fit and that is used instead (with values as tabulated in Table \ref{tab:xstar}). While no follow-up X-ray observations were obtained until $\sim$210 days after the discovery date, a fortuitously timed eROSITA scan \citep{2020ATel14246....1G} constrains the X-ray luminosity to $< 8 \times$10$^{42}$ erg s$^{-1}$ (3$\sigma$ upper limit corrected for galactic absorption and an additional extra-galactic column of n$_{\rm H, xgal}$ = 0.9$\times$10$^{20}$ cm$^{-2}$) at the UV/optical peak. This rules out bright X-ray emission around that epoch, and implies an X-ray brightening by a factor of $\gtrsim$25 over a 200 day period following the UV/optical peak. Early time faint X-ray emission at lower levels (below 8$\times$10$^{42}$ erg s$^{-1}$), observed for some other TDEs (e.g. \citealt{Gezari2017, Hinkle2021, Yao2022, Guolo23}), remains unconstrained. 

From the optical host galaxy spectrum (Section \ref{sec:data}), we measure a stellar velocity dispersion of $\sigma$ = 56$\pm$2 km s$^{-1}$. This translates into a black hole mass of log$_{10}$(M$_{\rm BH}$) = 5.2$\pm$0.46 M$_{\odot}$ using the M--$\sigma$ relation of \citet{Mcconnell13}, or alternatively log$_{10}$(M$_{\rm BH}$) = 6.1$\pm$0.35 M$_{\odot}$ using the \citet{Kormendy13} relation. This velocity dispersion (and hence black hole mass) is typical of other X-ray and UV/optical TDEs \citep{Wevers19}. We assume a value of log$_{10}$(M$_{\rm BH}$) = 6.1 $\pm$ 0.35 M$_{\odot}$ to calculate the Eddington ratio in the rest of this work. We note that this value is consistent with the black hole mass derived from the location of the inner disk radius, obtained from continuum X-ray spectral modeling (log$_{10}$(M$_{\rm BH}$) = 6.5 $\pm$ 0.6, \citealt{Mummery23}). 

\begin{table*}
    \centering
    \begin{tabular}{c|ccccccccccc}
     Instrument & Exp. time & Counts & Phase & MJD & Range & n$_{\rm H, xgal}$ & kT & L$_{\rm unabs}$ &L$_{\rm obs}$ & $\chi^2$ (dof) \\
     & (seconds) & & & (days) &(keV) & (10$^{20}$cm$^{-2}$) & (eV) & Log$\bigl($$\frac{\rm erg}{\rm s} \bigr)$ & Log$\bigl($$\frac{\rm erg}{\rm s} \bigr)$ & \\ \hline
       Swift/XRT & 32900& 666 & All & 59179 -- 59899 & 0.3--1.1 & 4$^{+5}_{-4}$ & 89 $\pm$ 9& 44.4$\pm$0.3 & 43.28$\pm$0.03 & 46 (60) \\
       NICER/XTI & 4412& 13779 & E12& 59187 -- 59189 & 0.3--1.1 & 0.9$\pm$0.3 & 110$\pm$3  & 44.54$\pm$0.01 & 43.76$\pm$0.02 & 18 (14) \\
      &5790 & 10419 & E34 & 59191 -- 59195 & 0.3--1.1 & 3.9$\pm$0.7 & 90$\pm$3 & 44.5$\pm$0.1 & 43.60$\pm$0.05 &  29 (14)\\
       XMM/PN & 21800 & 5255 & Late & 59725 & 0.2--0.9 & 1.8$\pm$1 & 63$\pm$3 & 44.35$\pm$0.15 & 42.93$\pm$0.05 &  22 (18) \\\hline
    \end{tabular}
    \caption{Continuum fitting results using a thermal (diskb blackbody) model including Galactic and extragalactic absorption ({\tt TBabs $\times$ zTBabs $\times$ zashift $\times$ clumin $\times$ diskbb}), which are used to generate the photo-ionization models. Range indicates the spectral range used for fitting. The intrinsic/unabsorbed luminosity L$_{\rm unabs}$ is quoted from 13.6 eV -- 13.6 keV, while L$_{\rm obs}$ is the observed/absorbed luminosity quoted for the range used in the spectral fitting.}
    \label{tab:continuumfits}
\end{table*}

\begin{table*}
    \centering
    \begin{tabular}{cc|ccccccccc}
    Epoch & Model & n$_{\rm H, xgal}$ & kT & Log(N$_{\rm H}$) & Log($\xi$) & Velocity & $\chi^2$ (dof) & $\Delta \chi^2$ (dof) & $\rm \Delta (AIC)$ \\
     &&(10$^{20}$ cm$^{-2}$)&(eV)& (cm$^{-2}$) &   & (c) & \\ \hline
     E12 & PCWA & 0.8$^{+0.8}_{-0.4}$ &114$^{+10}_{-5}$ & 22.1$^{\dagger}$ & 1.25$^{\dagger}$ & 0$^*$ & 16 (11) & 2 (3) & +4 \\
     E34 &  & 1.3$\pm$1 & 105$\pm$8 &23.1$^{+0.2}_{-0.8}$ & 2.15$\pm$0.8 & 0$^*$ & 11 (11) & 19 (3) & --17 \\
     XMM &  & 2.8$\pm$1& 60$^{+8}_{-3}$& 22.75$^{\dagger}$ & 0.75$^{+0.25}_{0.75}$& 0$^*$ & 19 (15) & 3 (3) & +3 \\
     E12 & UFO & $<$0.8 & 119$\pm$5 & 23.1$^{+0.7}_{-0.2}$ & 4.3$^{+0.6}_{-1.0}$ & --0.49$\pm$0.03 & 14 (11) & 4 (3) & +2 \\
     E34 &  & 1.3$\pm$0.9 &  108$\pm$6 & 22.7$^{+0.3}_{-0.6}$ & 3.4$^{+1.0}_{-0.8}$ & --0.14$\pm$0.03 &  4.5 (11) & 25 (3) & --21 \\
     XMM &  & 0.9$^{+2.8}_{-0.9}$ & 74$\pm$8 & 21.5$^{+0.6}_{-0.3}$ & 1.55$^{+1.0}_{-1.5}$ & --0.15$^{+0.05}_{-0.03}$ & 5 (15) & 17 (3) & --11 %\\
       \\\hline
     %E3 & 3451 & 7100 & 0.8$^*$ & 113$\pm$3 & 23.1$^{+0.05}_{-0.4}$ & 4.4$^{+0.2}_{-0.2}$ & --0.096$^{+0.03}_{-0.03}$ & 7 (9) & 22 (3)\\
     %E4$^{\dagger}$ & 2339 & 5050 & 0.8$^*$ & 110$\pm$8 & 22.2$^{+0.55}_{-0.15}$ & 3.0$^{+0.35}_{-0.35}$  & --0.14$\pm$0.03 & 4.6 (7) & 9 (3)\\\hline
    \end{tabular}
    \caption{Partial covering warm absorber (PCWA) and UFO fitting results. Uncertainties indicate the 90\% confidence intervals. For values marked with a $\dagger$ symbol the 90\% confidence range spans the entire parameter space. $\Delta \chi^2$ denotes the improvement in the fit statistic by adding a more complex component to the baseline {\tt diskbb} model (this denotes $\rm \Delta C-stat$ in the case of the XMM spectrum), where the additional number of degrees of freedom is denoted between brackets. The corresponding change in AIC is also noted. The velocity of the PCWA is fixed at 0 km s$^{-1}$ (indicated by an asterisk). The Galactic column density is fixed at 3.6$\times$10$^{20}$ cm$^{-2}$.}
    \label{tab:xstar}
\end{table*}

\subsection{X-ray spectral modeling}
\label{sec:specmod}
\begin{figure*}[!ht]
    \centering
    \includegraphics[width=0.48\textwidth]{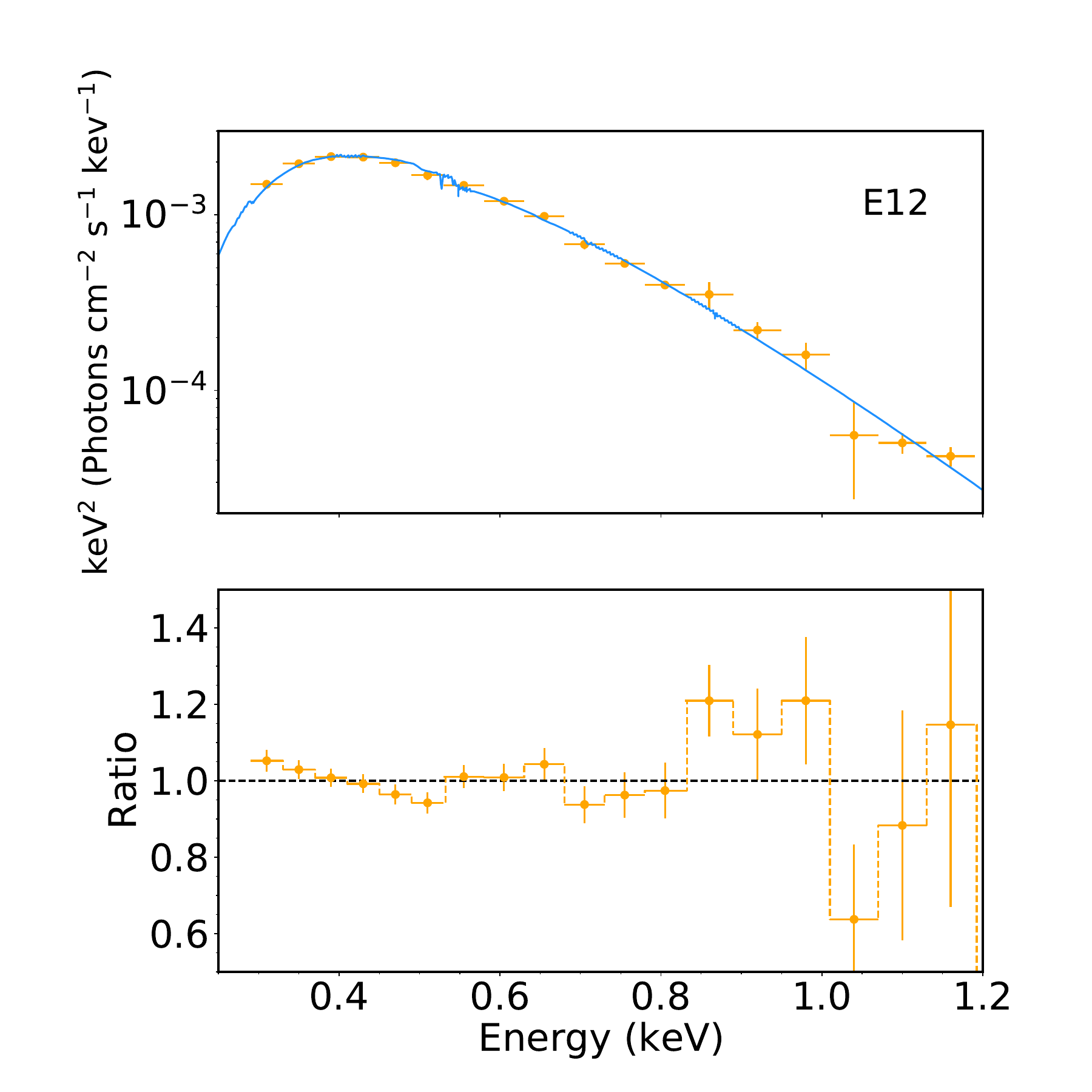}
    \includegraphics[width=0.48\textwidth]{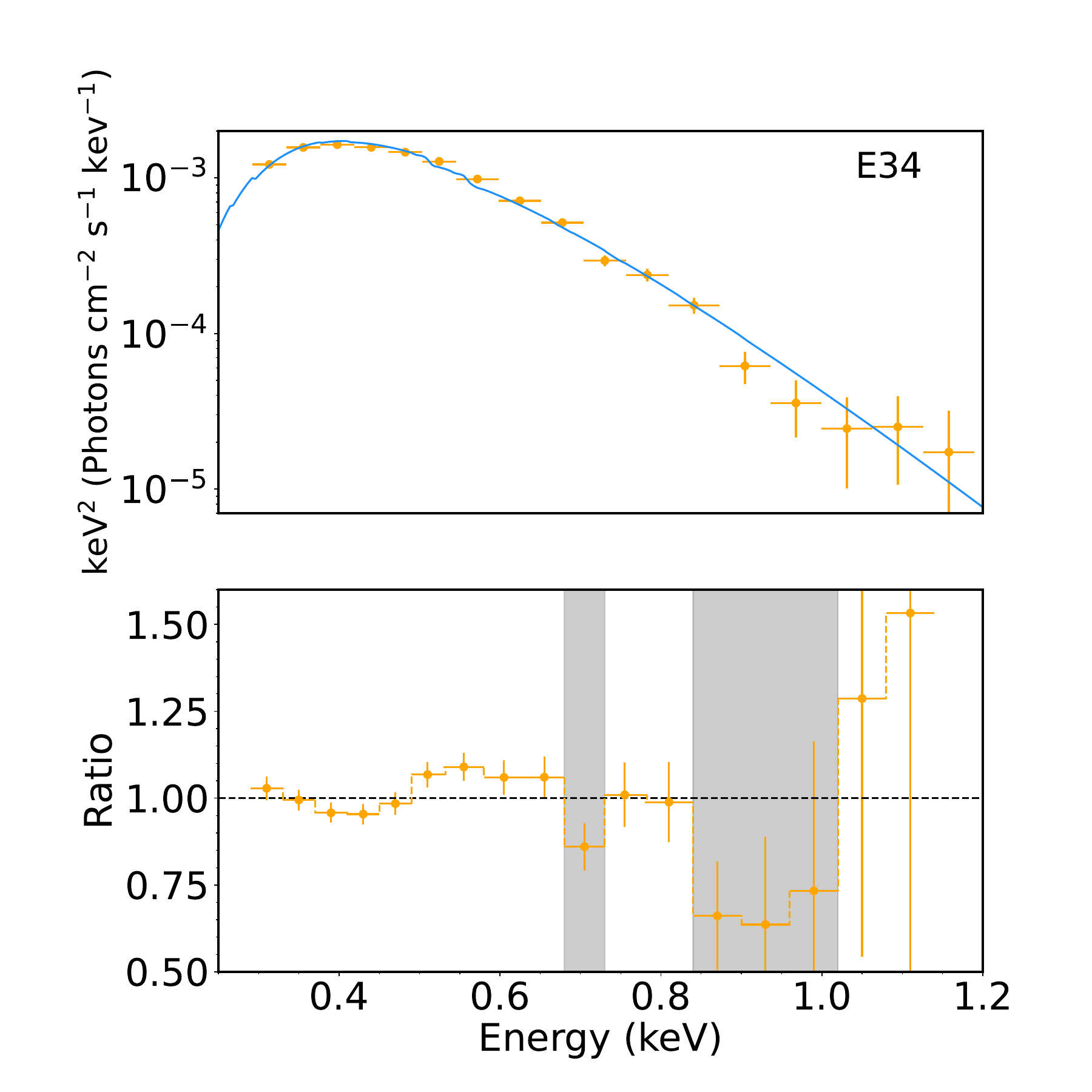}\\
    \includegraphics[width=0.48\textwidth]{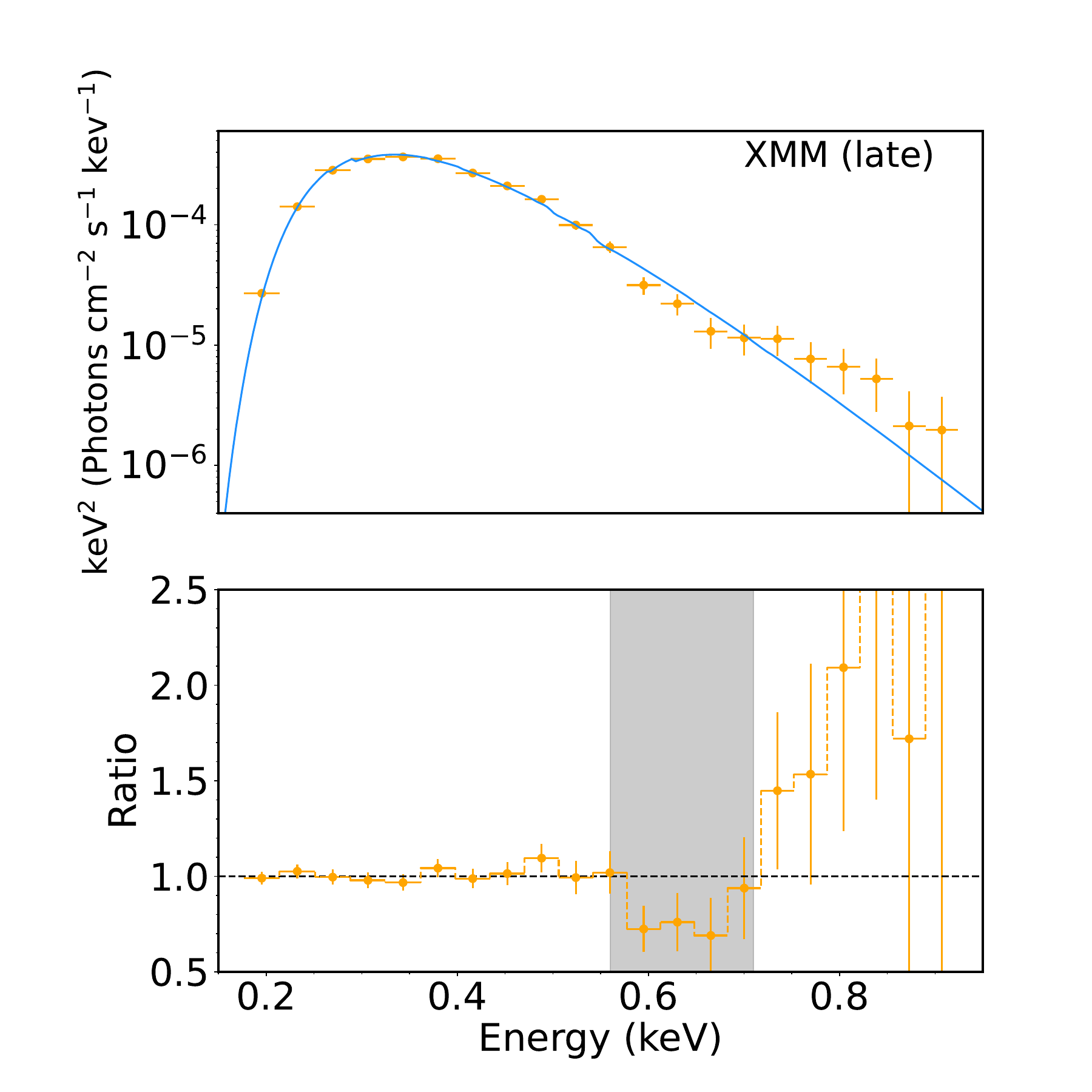}
    \caption{{\bf X-ray spectral fitting results.} Continuum fitting of the NICER and XMM-Newton X-ray spectra. Note the systematic residuals highlighted by the shaded areas. For an explanation of the difference in observed energy of the spectral residuals, see Section \ref{sec:uforesults}.} 
    \label{fig:xrayspec}
\end{figure*}

We analyse NICER X-ray data that are binned using the optimal binning scheme designed by \citet{kaastra} with the additional constraint of 20 counts per spectral bin. We accomplish this by using the options {\tt grouptype=optmin} and {\tt groupscale=20} in the {\it ftgrouppha} {\it ftool}. XMM-Newton data are binned to have at least 1 count per bin using the XMM data analysis tool {\it specgroup} with {\it oversample=3}. We require a higher number of counts per bin for the NICER data to avoid uncertainties introduced by the empirically estimated background \citep{3c50}. For XMM-Newton this component is much better constrained directly from the observations, allowing robust data analysis even with a lower number of photons per bin. As a result, for the spectral fitting we employ $\chi^2$ statistics for the NICER data and Cash statistics \citep{cstat} for the XMM-Newton spectrum, using the {\tt Xspec} software package \citep{xspec} distributed with HEASoft version 6.32. We note that using Cash statistics for the NICER spectral fitting does not change the results.

\subsubsection{Continuum fitting: systematic residuals}
We model the spectra with an absorbed thermal (disk blackbody\footnote{For simplicity we use the standard {\tt diskbb} model rather than a more complex thermal model with more free parameters, such as that presented in \citealt{Mummery23}.}) continuum model ({\tt TBabs $\times$ zTBabs $\times$ zashift $\times$ diskbb}).  The Galactic Hydrogen column density n$_{\rm H}$ is fixed to 3.6$\times$10$^{20}$ cm$^{-2}$ \citep{hi4pi}, while the extragalactic value n$_{\rm H, xgal}$ is left as a free parameter. We also calculate the {\it bolometric} ionizing luminosity by integrating this model from 1--1000 Rydberg (using the {\tt clumin} model in {\tt Xspec}); we find L$_{\rm ion} \approx 3 \times 10^{44}\ \rm erg\ s^{-1} \approx 2$ L$_{\rm Edd}$, which remains steady even 770 days after UV/optical peak (Table \ref{tab:continuumfits}). This can be understood by taking into account the temperature (among other factors) dependent bolometric correction, shifting part of the X-ray emission out of the observed band \citep{Mummery23}, but we caution that the extrapolations involved are far out of the observed band and hence very sensitive to the model assumptions. 
Because these luminosities are very sensitive to the amount of extinction, which varies for different spectral fits, we have also included a column which lists the uncorrected, i.e. absorbed, luminosity for each epoch in Table \ref{tab:continuumfits}.

The results of the continuum fitting are presented in Table \ref{tab:continuumfits}. For the second NICER epoch E34, the high $\chi^2$ values are caused by systematic residuals, which are evident in Figure \ref{fig:xrayspec}. For the XMM-Newton spectrum, very similar systematic residuals are present above 0.5 keV. We have carefully checked that spectral residuals are not dependent on the background region chosen, and are independent of the details of the data binning, the source extraction region and the exact energy range used for fitting. We have also investigated the possibility of background contamination for the NICER data, described in detail in Section \ref{sec:nicer}.
Similar systematic residuals have been observed in other highly accretion black hole systems (including TDEs), and explained by blueshifted absorption features of highly ionized species including O\,\textsc{vii} (E = 0.739 keV) or O\,\textsc{viii} (E = 0.653 keV) within the line of sight. 

\subsubsection{Modeling the residuals: power-law and partial covering warm absorber (PCWA) do not work well}
\begin{figure}
    \centering
    \includegraphics[width=0.25\textwidth]{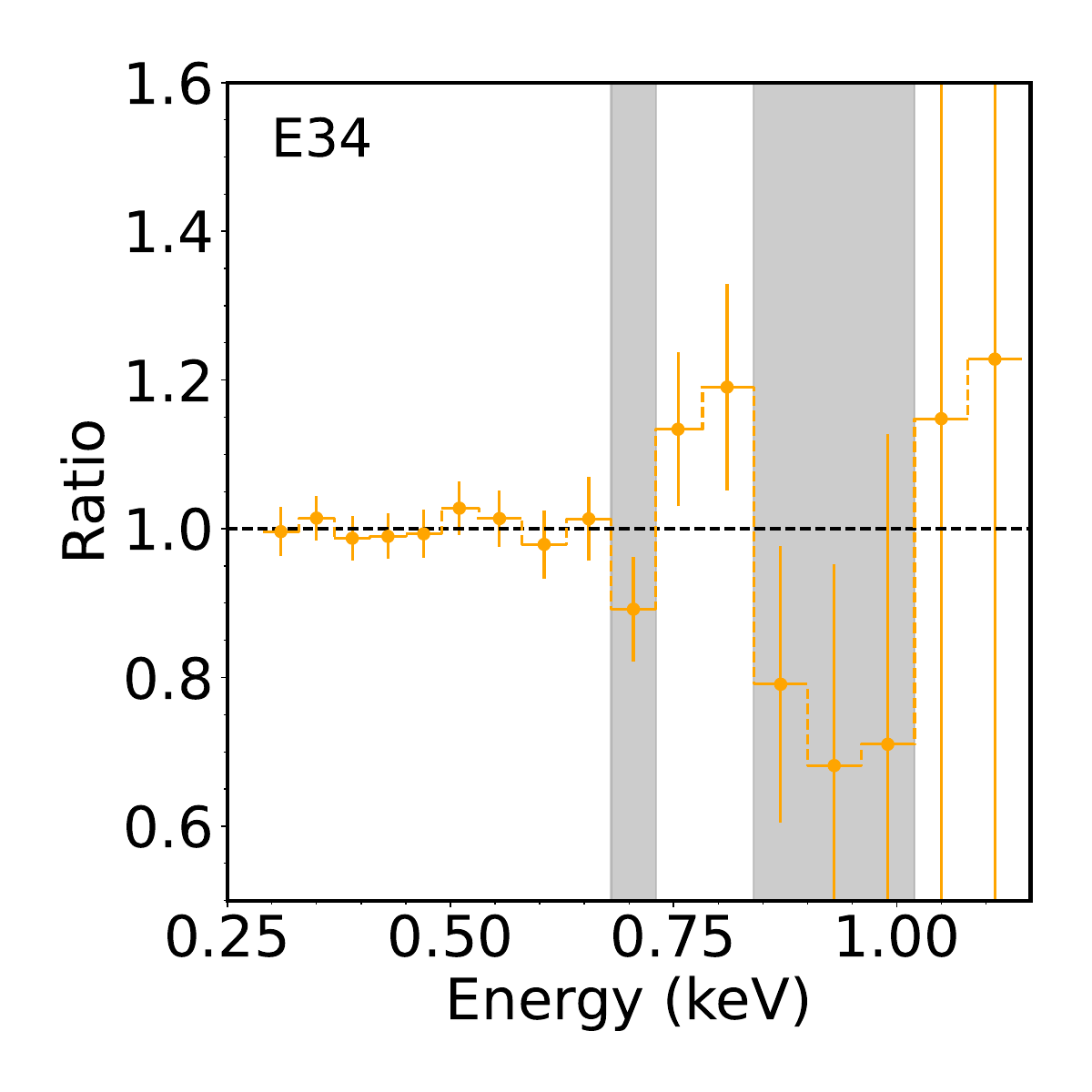}\includegraphics[width=0.25\textwidth]{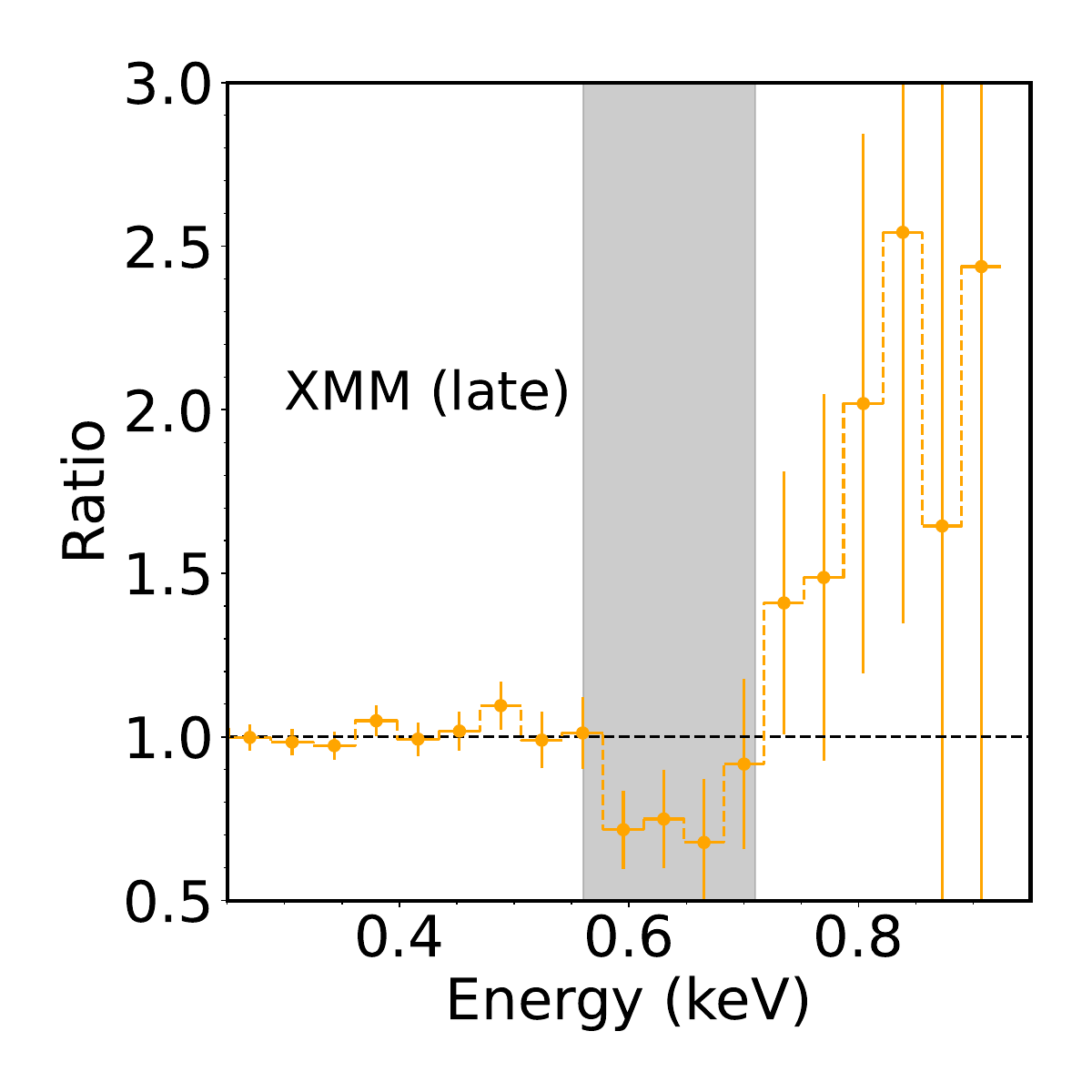}\\
    \includegraphics[width=0.25\textwidth]{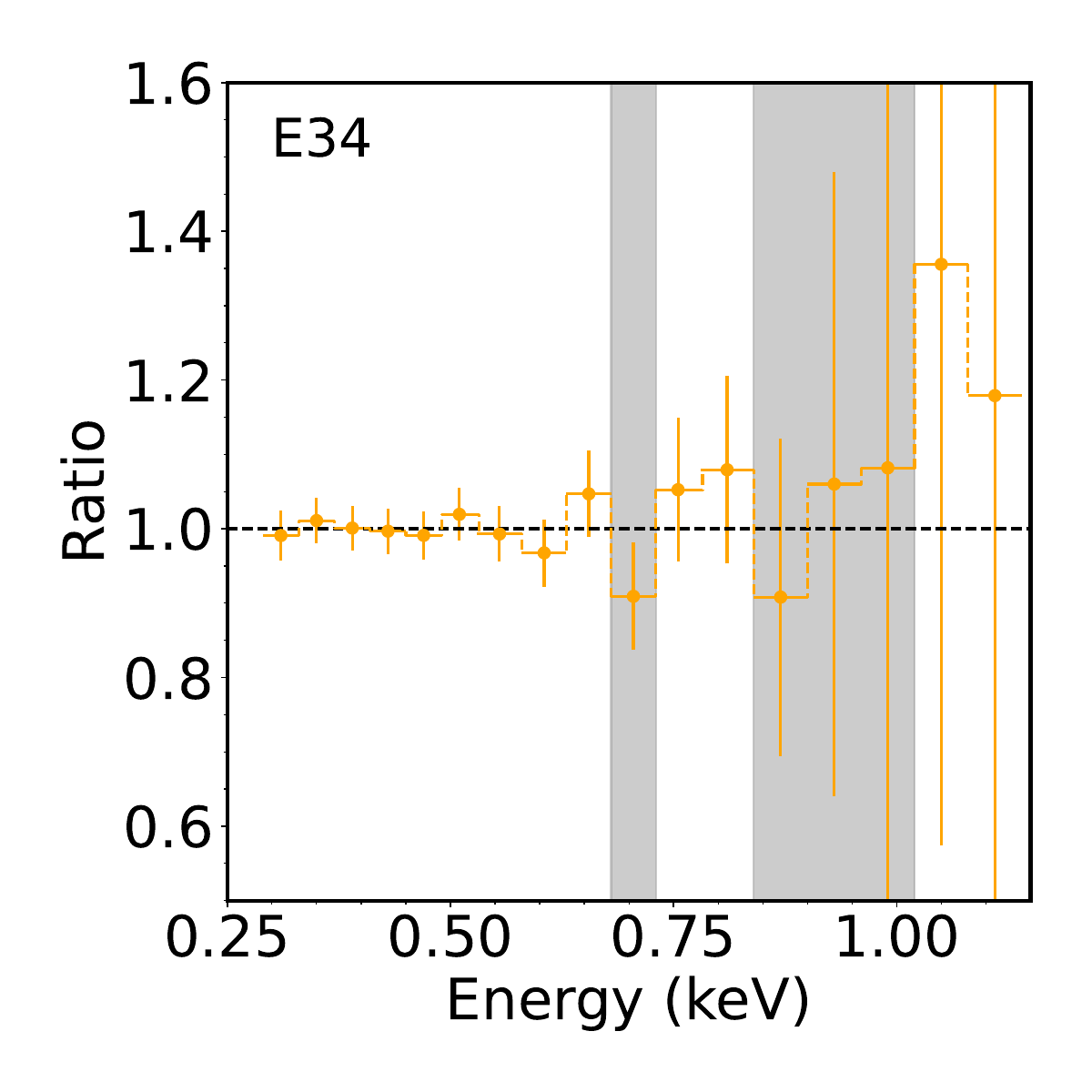}\includegraphics[width=0.25\textwidth]{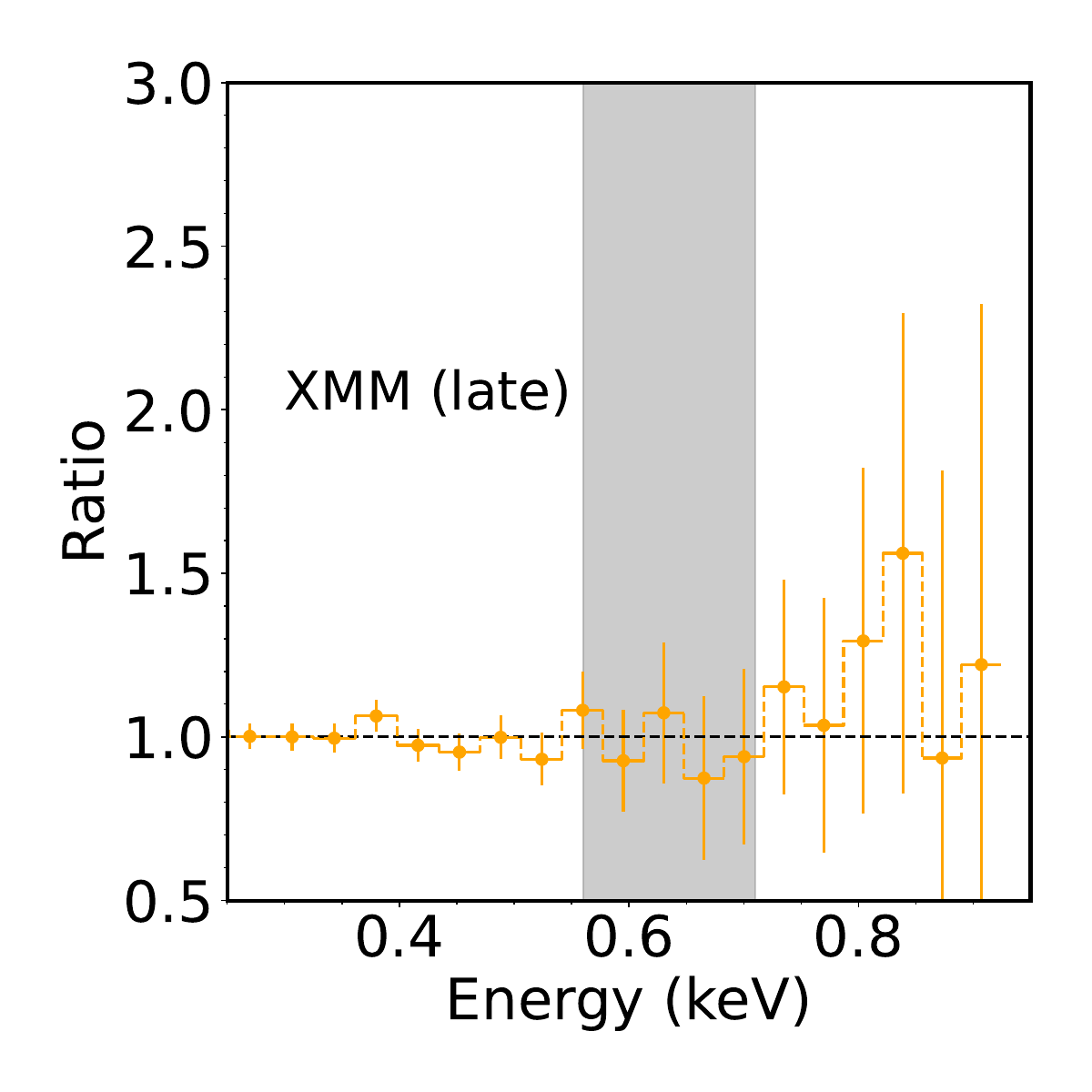}
    \caption{Side-by-side comparison of the ratio between the data and the best fit for the continuum+PCWA (top) and continuum+UFO (bottom) models for epochs E34 and XMM. The systematic residuals  remain in the PCWA model (top panels), but disappear in the UFO model. The energy ranges where the absorption features occur are highlighted with a grey band. Scales are matched for the E34 and XMM panels for easy comparison. It is clear that the UFO model better describes the data than the PCWA model.}
    \label{fig:ratio}
\end{figure}

The presence of a (weak) power-law or an additional thermal component can mimic the residuals seen in Figure \ref{fig:xrayspec}. We therefore first investigate whether either of these models, in combination with the continuum {\tt diskbb} model, can improve the fits. Adding a power-law, single temperature ({\tt bbody}) or multi-temperature ({\tt diskbb}) blackbody model or a brehmstrahlung model does not improve the fit statistic\footnote{Combining either of these models with a brehmstrahlung model (e.g. {\tt bremss+powerlaw} instead of the baseline {\tt diskbb} model) does not improve the fit statistic.}. None of these models is able to satisfactorily describe the combination of the absorption feature and the higher continuum level at higher energies. Specifically, adding a power law component and leaving the index $\Gamma$ free results in a negative best-fit index $\Gamma$ = --2.5 (which is unphysical), and the normalization becomes very small (of order 10$^{-12}$). If we force $\Gamma$ to be positive, it tends to very high values (up to $\Gamma \approx$ 9, which is again unphysical) and the normalization tends to small values $<$ 10$^{-7}$. In other words, the power law component contribution is negligible, not statistically required, and the spectral residuals look similar to those of a thermal-only model fit. When adding another thermal {\tt diskbb} component, either i) the temperature of the second component is identical to that of a single thermal model fit, but the normalization doubles, resulting in the same continuum shape for the model combination, or ii) the temperature of the second component becomes very small (of order 0.0002 keV) and the normalization very large, leading to a negligible contribution to the total continuum shape. Again, the resulting continuum shape (and residuals) are similar to that of a single thermal component fit. Next, we investigate whether a PCWA (i.e. partial covering, low velocity ionized gas along our line of sight) can provide an improvement to the fit. To this end we generate Xstar photo-ionization models with a low (100 km $s^{-1}$) intrinsic velocity broadening. 
To approximate the ionizing continuum we use the best fit temperature from the thermal model and the appropriate ionizing luminosity. We include a wide grid of parameter ranges to avoid finding best solutions near the range edges.
The results of this modeling can be found in Table \ref{tab:xstar}. In summary, the PCWA model does provide an improvement in the fit statistic for the NICER ($\Delta \chi^2$ = 2 and 19 for epochs E12 and E34, for 3 additional degrees of freedom) and XMM-Newton ($\Delta C$-stat = 3 for 3 additional degrees of freedom) spectra. However, the required column densities are on the high end (between 10$^{22}$ and 3$\times$10$^{23}$ cm$^{-2}$), systematic residuals still remain (Figure \ref{fig:ratio}), and as discussed in more detail in Section \ref{sec:aic}, the addition of a PCWA model is statistically not warranted for the XMM-Newton spectrum. 

\subsubsection{Modeling the residuals: ultra-fast outflow}
\label{sec:uforesults}
As an alternative model to explain the residuals, based on theoretical expectations for the post-disruption evolution of the debris (e.g. \citealt{Dai2021}), we also consider an ultra-fast outflow model. This model was used by \citet{Kara18} to model broad, systematic residuals in the TDE ASASSN-14li.
We generate photo-ionization models using XSTAR with a high velocity broadening (30\,000 km s$^{-1}$, the maximum available in {\tt xstar}). Such components are designated as ultra-fast outflows if their velocity shift exceeds 10\,000 km s$^{-1}$ \citep{Tombesi10}, to discriminate them from lower velocity components in multi-phase wind systems (see e.g. \citealt{Kosec23} for an example in the TDE ASASSN--20qc). % 

We incorporate the ionized high velocity absorption model into the previously described continuum model, adding its equivalent Hydrogen column density (N$_{\rm H}$), ionization parameter ($\xi$) and velocity shift (v${\rm _w}$) as free parameters to the fit. The results are presented in Table \ref{tab:xstar}. For spectrum E12 the improvement in the fit statistic is $\Delta \chi^2$ = 4, i.e. not statistically significant. We find a very high velocity ($\approx -0.49$ c), and a high ionization parameter log($\xi$) = 4.3$^{+0.6}_{-1.0}$ and column density N$_{\rm H}$ = 13$^{+50}_{-5} \times 10^{22}$ cm$^{-2}$. 

For spectrum E34, this model leads to an improvement in the fit statistic of $\Delta \chi^2$ = 25 for 3 degrees of freedom, equivalent to a p-value of $<10^{-5}$, or an F-test probability of 0.000475. We further verified this result by running 100\,000 simulations of the spectrum and calculating the difference in fit statistic when adding the ionized absorber using the {\it simftest} routine in {\tt Xspec}. We found 4 occurrences where the fit statistic improvement was equal to or greater than that of the observed data, implying a probability of this improvement being due to statistical noise of $\sim $10$^{-4}$ after accounting for the number of spectra (3) that were searched. This is roughly consistent with the F-test results. 

For E34, we find values of v$_{\rm w}$ = --42\,000$\pm$9\,000 km s$^{-1}$ (--0.14$\pm$0.03c), log($\xi$) = 3.4$^{+1.0}_{-0.4}$ and N$_{\rm H}$ = 5$\pm$4$\times 10^{22}$ cm$^{-2}$. There are no more systematic trends in the residuals after fitting (Figure \ref{fig:ratio}). These values are consistent with an origin of the absorption feature due to a range of transitions including the O\,\textsc{viii} doublet at a rest-frame energy of 0.65 keV, as well as Fe\,\textsc{xvii} and Ne\,\textsc{ix} (see Figure \ref{fig:ionline} in the supplementary materials). Modeling the individual epochs E3 and E4 yields results that are consistent with the combined spectrum E34, but at lower significance, implying that there is no significant variability on shorter timescales (see Section \ref{sec:intravar} for more details).

For the XMM-Newton spectrum, the improvement in the fit statistic is $\Delta C$-stat = 17. Performing the same procedure as described before, we find 8 occurrences out of 25\,000 simulations where the fit statistic improvement is equal to or greater than that observed in the data (probability of $\sim $10$^{-3}$ after accounting for the number of spectra searched) We find a velocity of v$_w$ = --0.15$^{+0.05}_{-0.03}$ c, while log($\xi$) = 1.55$^{+1.0}_{-1.5}$ and N$_{\rm H}$ = 3$^{+8}_{-1.5} \times$ 10$^{21}$ cm$^{-2}$ are lower by factors of 100 and 10 respectively at late times (note the logarithmic in $\xi$). Given the decrease in continuum (disk blackbody) temperature by a factor of almost 2 compared to E34, a plausible explanation for this evolution is a corresponding change in the ionization balance and optical depth of the ionic transitions causing the absorption feature. This can also help explain the change in the energy at which the feature appears (see Fig. \ref{fig:ionline}), despite the best fit UFO model requiring a near constant wind velocity.

We derive an upper limit on the column density in epoch E12 by assuming the best fit parameters of the detection in E34; we find N$_{\rm H}$ $< 1.1 \times 10^{21}$ cm$^{-2}$. 

\subsubsection{Model selection: Akaike information criterion}
\label{sec:aic}
Both the PCWA and UFO models provide an improvement in the fit statistic of the spectra in E34 and XMM (Table \ref{tab:xstar}). 
Model selection between simple baseline and increasingly complex models can be done through a number of metrics. Here we employ the Akaike information criterion (AIC; \citealt{Akaike1974}) as a means to ascertain whether the addition of a complex component (in addition to the basic disk blackbody model) is warranted. We compute the difference in AIC as \begin{equation} 
\rm \Delta(AIC) = \Delta(\chi^2\ or\ C stat) + 2 \times \Delta(dof)
\end{equation} where $\rm \Delta(dof)$ is the additional number of parameters introduced by the model. $\rm \Delta(dof)$ = 3 for the PCWA and the UFO models\footnote{The line broadening is an additional free parameter for both the PCWA and UFO models, although it was fixed at 100 and 30\,000 km s$^{-1}$, respectively. Taking into account this extra degree of freedom does not change any of the results.}. An improvement in $\rm \Delta(AIC)$\footnote{A decrease in fit statistic leads to a decrease in AIC, hence a negative sign indicates an improvement.} of better than --5 (--10) signifies a strong (very strong) preference for a more complex model. By rank ordering several complex models, the same guidelines can be used to decide which of these may be preferred by the data, if any (i.e. if the $\rm \Delta(AIC)$ between 2 complex models is larger than --10, e.g. \citealt{deltaaic}). 

We start by noting that the $\rm \Delta(AIC)$ is positive for E12 for both models, indicating that adding a more complex model is not statistically favored by the data.
Because it is implausible that the absorption feature is caused by the PCWA in E34 and the UFO in the XMM spectrum (or vice versa), we consider the joint $\rm \Delta(AIC)$ for those 2 observations.
From the values in Table \ref{tab:xstar}, we see that $\rm \Delta(AIC)$ = --14 for the PCWA model, and $\rm \Delta(AIC)$ = --32 for the UFO model. This implies that statistically speaking, the UFO model is strongly preferred over the PCWA model in the assumption that the feature in both spectra has the same physical origin. 

The UFO model provides a natural link between the TDE and the high velocity absorber, as it is expected that a powerful wind is launched at the expected high Eddington accretion rates \citep{Rees1988}. In addition to being preferred in a purely statistical sense (as inferred from the AIC analysis), this scenario has the added benefit of providing a natural explanation for the almost identical properties of AT2020ksf and ASASSN-14li described in Section \ref{sec:14li}.

Although the analysis described here suggests that the UFO model is statistically preferred, it remains possible that the underlying continuum may be much more complex than a pure thermal component. However, for most other TDEs a thermal component describes the continuum well \citep{Guolo23}.

In light of these results -- that the UFO model provides the statistically preferred description of the early+late time data -- we assume that an ultra-fast outflow is present following the TDE from E34 onwards, with properties as tabulated in Table \ref{tab:xstar}.

\subsection{Outflow energetics}
Assuming that the wind is launched at the point where its velocity surpasses the escape velocity of the black hole, we infer that the launching radius is located around $r_l = 100 \rm R_{\rm g}$, where $\rm R_{\rm g}$ = $\frac{\rm 2 G M_{\rm BH}}{c^2}$ is the gravitational radius of the black hole.
Instantaneous quantities such as the mass outflow rate can then be estimated as \citep{Nardini2015} $\dot{M}_{\rm out} = \rm \Omega m_p N_H v_w r_l $, where $\Omega$ is the covering factor of the wind (assumed to be 2$\pi$ for simplicity), m$_{\rm p}$ is the proton mass, N$_{\rm H}$ is the column density,  and v$_w$ is the wind velocity. Taking the best-fit parameters for E34, this leads to
\begin{equation}
    \dot{M}_{\rm out} \approx 8 \times 10^{22} \rm g\ s^{-1} = 0.048 \dot{M}_{\rm Edd}
\end{equation} 
or $\sim$0.001 M$_{\odot}$ yr$^{-1}$ (for XMM this is $\sim 7.5 \times 10^{-5}$ M$_{\odot}$ yr$^{-1}$).
A strict upper limit for the wind launching radius can be derived by taking into account the variability timescale. The shortest time between the initial non-detection and the UFO detection (i.e. the end of E12 and the beginning of E3) is $\sim$2 days; this corresponds to a light travel distance of 5$\times$10$^{15}$ cm $\approx$ 14\,000 R$_{\rm g}$, which translates to $\dot{M}_{\rm out} \lesssim 7\dot{M}_{\rm Edd}$. These limits are consistent with the observed Eddington ratio of $\sim$1, and similar to typical values derived analytically (e.g. \citealt{Thomsen22}).

The (instantaneous) kinetic power of the wind can now be estimated as 
\begin{equation}
    \rm P_{\rm kin} = \frac{1}{2} \dot{M}_{\rm out} v^{2}_{w} \approx 7 \times 10^{41} erg\ s^{-1} = 0.004 \rm L_{\rm Edd}
\end{equation}
for E34; the upper limit derived from the variability timescale translates to P$_{\rm kin} \lesssim 0.56 \rm L_{\rm Edd}$. Compared to the bolometric radiative output of the system, this represents 0.002L$_{\rm bol}$ $<\ $P$_{\rm kin} <$ 0.3L$_{\rm bol}$. 

The total energy injection can be estimated by assuming that the UFO column density does not exceed the upper limit found for E12 in the period starting from the discovery date, and that the UFO has properties as calculated in Table \ref{tab:xstar} during its further evolution. For simplicity we use the E34 wind properties until the midpoint with the XMM-Newton observation (i.e. MJD 59360), and the properties of the latter afterwards. In this way, we find that E$_{\rm kin} \sim 2 \times 10^{49}$ erg. 

Comparing this to the galaxy bulge gravitational binding energy, E$_{\rm bulge} \sim \rm M_{\rm bulge} \sigma^2 \sim 10^{54}$ erg, indicates that the energy input of a single TDE UFO is negligible. We will return to the cumulative effect of TDEs in more detail in Section \ref{sec:feedback}.

\section{Discussion}
\label{sec:discussion}
We have established that a PCWA model can describe the absorption features well in only one of the two instances where they are observed. An ionized, fast-moving absorber provides a good model description of the absorption features for both epochs. We now assume that the absorption features represent an ionized, ultra-fast outflow or disk wind. 

\subsection{Ultra-fast outflow variability}
While the absorber is not statistically required in the epoch E12, it is significant in epochs E34 and XMM. 
We find moderate velocities ($\approx 0.15$c), and a high but decreasing ionization parameter (log($\xi$) = 3.4 to log($\xi$) = 1.5) and column density (from 1--5$\times$10$^{22}$ cm$^{-2}$ to 3$\times$ 10$^{21}$ cm$^{-2}$). 

To further establish this result, we have investigated possible scenarios to explain the changes in the X-ray spectrum without a UFO for the NICER data, specifically between E12 and E34. First, the continuum emission for these epochs can be fit with a disk blackbody model with similar temperatures and luminosities (Table \ref{tab:continuumfits}). The observed X-ray luminosity is $\sim$25\% lower in E34 compared to E12, while the bolometrically corrected disk luminosity is similar in both epochs. This allows us to rule out continuum variability as the origin of the difference. Second, both spectra contain a similar number of X-ray photons ($>$10\,000), so SNR related issues can also be discarded. Third, we have carefully investigated the potential contribution of a telluric Oxygen emission line to the NICER spectra (see Section \ref{sec:nicer} for details), and rule out that this significantly affects the spectra. 

We can therefore robustly establish that the UFO properties were rapidly changing on a timescale of days. Such rapid variability of the disk wind has not been reported for another TDE. The UFO observed in ASASSN--14li was reported to be stable over $\sim$year timescales \citep{Kara18}; however, more recently Ajay et al. (submitted) report variability on a timescale of months (see Section \ref{sec:14li} for a more detailed comparison). Variability in UFOs has been seen in high accretion rate stellar mass compact objects (e.g. ultra-luminous X-ray sources, \citealt{Kosec2018}), and interpreted as a sign of a clumpy wind on the basis of MHD simulations \citep{Kobayashi2018}. UFOs with similar properties have also been observed in AGNs, with a range of velocities and ionization parameters (see e.g. \citealt{Tombesi10} and references there-in). Variability has been seen on a range of timescales, going from years down to less than a day \citep{Cappi06, Turner2007, Saez2009, Pinto2021}. The short timescales involved here, coupled with the lack of significant luminosity/continuum evolution, make it unlikely that scenarios such as the over-ionization of the wind material (e.g. \citealt{Pinto2018}) or variable self-screening effects \citep{Pinto2020} can explain the observed rapid variability.

Short timescale variability has also been attributed to changes in the column density, opacity and covering fraction in the framework of multi-phase and/or clumpy winds (e.g. multiple components launched at different radii and existing in different ionization states depending on the direct line of sight to the central ionizing source; e.g. \citealt{Tombesi15}).
In this scenario, the approximate timescale for variability depends on the size of the clumps, their distance from the black hole as well as the black hole mass; simulations of supercritical accretion flows find variability timescales of $\sim$ 3 days for a 10$^6$ M$_{\rm BH}$ for typical wind/clump parameters \citep{Kobayashi2018}. This is well matched to the timescale of variability observed in AT2020ksf. 

\subsection{Comparison to ASASSN--14li}
\label{sec:14li}
AT2020ksf is the second TDE (after ASASSN--14li, \citealt{Kara18}) for which absorption signatures indicative of an ultra-fast outflow are detected in the X-ray spectrum. The similarities between these two sources are striking, and can be summarized as follows: 
\begin{enumerate}
    \item They display bright early-time UV emission, which reverts to a plateau on a timescale of $\sim$200 days.
    \item They have thermal X-ray emission which peaks around the Eddington limit, and gradually cools over time. While the observed luminosity decreases, when correcting for out of band emission the X-ray luminosity remains L$_X \sim$L$_{\rm Edd}$ for more than 770 days.
    \item No corona forms in either source even after 770 days, contrary to other TDEs with long-lived X-ray emission \citep{Wevers2021, Yao2022}. This is consistent with the high Eddington fraction $\gtrsim 0.1$ f$_{\rm Edd}$ \citep{Wevers20}.
    \item The inferred black hole masses are similarly $\sim$10$^6$ M$_{\odot}$ \citep{Miller2015, Wevers17}.
    \item Both sources show UFO signatures in their X-ray spectra, even more than 2 years after UV/optical peak (as shown by Ajay et al. submitted).
    \item Both sources are detected at radio wavelengths, and have similar radio luminosities at their respective phases ($\sim$2-6$\times$ 10$^{37}$ erg cm$^{2}$ s$^{-1}$ at 6 GHz, roughly 200 days after optical peak, \citealt{Bright18}).
\end{enumerate}
These similarities suggest that they may represent {\it twin} systems where the same physical conditions are prevalent. We note that AT2020ksf is a factor of $\sim$5 further away in distance than ASASSN--14li.\\
Significant differences are found in two respects: i) the integrated column density of the UFO, which is higher by a factor of $\sim$10 in AT2020ksf at early times, and ii) the decoupling between the UV/optical and X-ray components. 
The former could suggest that the outflow has a much higher mass loading factor. As we have discussed previously, however, UFO variability complicates this interpretation. 
We discuss possible explanations for the difference between ASASSN--14li, which was promptly X-ray bright, and AT2020ksf, which showed a 230 day X-ray delay, in the next section.

\subsection{The delay between the UV/optical and X-ray peaks}
\label{sec:delay}
The UFO model parameters indicate a rapidly varying column density. Other notable properties of AT2020ksf are the detection of faint radio emission around the X-ray peak, and the $\sim$230 day delay between the UV/optical and X-ray lightcurve peaks. 

Several explanations have been proposed in the literature for the X-ray delay, including the clearing of an optically thick Eddington envelope which can be either neutral or ionized, and complete or partially covering (e.g. \citealt{Kajava2020, Guolo23}), as well as a pressure supported Eddington envelope \citep{Metzger22}, and the delayed onset of accretion (i.e. inefficient circularization, e.g. \citealt{Piran15, Gezari2017}). 

One scenario is that the stream-stream intersection promptly led to the formation of an accretion flow around the black hole, which then powers the optical/UV lightcurve. 
No large excess in neutral absorption is seen for AT2020ksf; to explain the non-detections a column of n$_{\rm H} \gtrsim 6 \times 10^{21}$ cm$^{-2}$ is required. This neutral material would then need to clear out completely after $\sim$200 days, when very little excess neutral column is required by the X-ray spectral fits after accounting for an ionized absorber (Table \ref{tab:xstar}). This material could in principle be expelled in a collision-induced outflow, which can have a significant (but less than unity) covering fraction \citep{Lu2020}. This would require the high optical depth portion of this component to block our line of sight to explain the X-ray delay for AT2020ksf, while in ASASSN--14li it would remain completely out of our line of sight. We note that hydrodynamics simulations predict outflow velocities $<$0.1c for a 10$^6$ M$_{\odot}$ black hole \citep{Lu2020}, which is inconsistent with the observed UFO velocities, and further that similar outflows have not yet been seen in global simulations \citep{steinberg, fangyi}.

Alternatively, the fallback rate in a TDE (i.e., the rate at which disrupted stellar material is supplied to the black hole) following the peak is approximately
\begin{equation}
    \dot{M} = \dot{M}_{\rm p}\left(t/t_{\rm p}\right)^{-\alpha}
\end{equation}
with \citep{coughlin22}
\begin{equation}
    \dot{M}_{\rm p} = \frac{M_{\star}}{4 t_{\rm p}}, \quad t_{\rm p} = t_0 M_{\rm BH, 6}^{1/2}, \label{tpeak}
\end{equation}
where $M_{\star}$ is the mass of the disrupted star, $M_{\rm BH,6}$ is the black hole mass in units of $10^6 M_{\odot}$, and $t_0 \simeq 25$ d is largely independent of the stellar properties \citep{ryu2020, bandopadhyay23}. If the star is completely destroyed, then $\alpha = 5/3$ \citep{Rees1988}, while if it is partially destroyed\footnote{However, if a large fraction of the star survives the encounter, there is an additional dependence on the pericenter distance that is not captured in Equation \ref{tpeak} \citep{guillochon13}. Therefore, these expressions are only valid if a substantial fraction of the stellar envelope is stripped during the tidal encounter with the star; see, e.g., Figure 1 of \citet{nixon21}}, $\alpha = 9/4$ although there is likely some evolution over time \citep{guillochon13, coughlin19, miles2020}. If the luminosity associated with the fallback accretion is $L = \eta \dot{M} c^2$, where $\eta \simeq 0.1$ is the radiative efficiency, and we equate this luminosity to $\ell\times L_{\rm Edd}$, where $L_{\rm Edd} = 4\pi GcM_{\rm BH}/\kappa$ is the Eddington luminosity with $\kappa \simeq 0.34$ cm$^2$ g$^{-1}$ the electron scattering opacity (assuming typical abundances), then the time-dependent Eddington fraction $\ell(t)$ is
\begin{equation}
    \ell(t) = 140 M_{\star,\odot} M_{\rm BH,6}^{\alpha/2-3/2}\left(t/t_0\right)^{-\alpha}, \label{ell}
\end{equation}
where $M_{\star,\odot}$ is the mass of the star in solar masses. Setting Equation \ref{ell} equal to 1 then yields the time at which the accretion luminosity falls below Eddington:
\begin{equation}
    \frac{t_{\rm Edd}}{t_0} = \left(140 M_{\star,\odot}\right)^{1/\alpha}M_{\rm BH,6}^{1/2-3/(2\alpha)}.
\end{equation}
With $\alpha = 5/3$, $M_{\star,\odot} = 1$, and $M_{\rm BH,6} = 1$, the previous expressions give a peak Eddington fraction of $\ell_{\rm peak} = 140$ and $t_{\rm Edd} \simeq 500$ d. During the super-Eddington phase, the accretion flow should be highly geometrically and optically thick and the viscous timescale (from the tidal radius) comparable to the dynamical time, being $\sim few$ hours (see Equation 3 in \citealt{dexter19}); the latter provides a self-consistency check on the assumption that the fallback and accretion rates are highly coupled.

It therefore seems plausible that the supercritical fallback rate at early times inflated the accretion flow into a quasi-spherical envelope. If the line of sight to the source is off-axis (with respect to the rotational axis of the accretion flow), the envelope simultaneously obscures both the X-ray emission and the outflow generated by the radiation pressure from the super-Eddington accretion (e.g., \citealt{roth16, Dai2018,Thomsen22}). 
For the X-ray emission in AT2020ksf to remain undetectable (at the level of the eROSITA non-detection) at early times would require an ionized column of N$_{H} \approx 8.5 \times 10^{23}$ cm$^{-2}$ (a factor 17 higher than observed in E34, assuming that all wind parameters remain the same). As time advances and the Eddington fraction drops, the scale height of the disc declines, allowing us to peer farther into the outflow (i.e., the
X-ray emission and wind are less obscured). As the accretion rate drops below Eddington the decreasing flux from the radiation field drives an outflow decreasing in power. In this model, therefore, the nearly contemporaneous emergence of the X-rays, the appearance of the UFO, and the decrease in power of the UFO are a direct byproduct of the accretion rate falling below the Eddington limit of the black hole, the timescale for which is in good agreement with observations if the disruption was of a solar-type star and complete. 

Another alternative is that the optical/UV emission is powered by the stream-stream collision following the periapsis advance of the returning debris stream \citep{Piran15}, and the formation of the smaller-scale (i.e., nearer the black hole) accretion disc powering the X-ray emission and launching the UFO was delayed by $\gtrsim 200$ days. In this case, since the stream-stream collision is powering the optical/UV emission, the delayed onset of X-ray emission cannot be generated by Lens-Thirring precession and, thus, the non-self-intersection of the debris stream \citep{Kochanek94, guillochon15}.
%, and must instead be due to an inherently long viscous delay. Not only is there little to no observational evidence of such viscous delays \citep{mockler19, nicholl22}, %(i also think that since the observed tde rate is ~ consistent with theoretical predictions, you can't use the argument that we wouldn't see them if they're viscously delayed
In the assumption of a compact, circular disk this means that the coefficient of dynamic viscosity (or the $\alpha$-viscosity in the case of a Shakura-Sunyaev disc; \citealt{shakura73}) would have to be extraordinarily small to explain a delay of $\sim 200$ days if the stream angular momentum establishes the circularization radius of $\sim 2 r_{\rm p} \simeq 40 r_{\rm g}$, where $r_{\rm g} = GM/c^2$ is the gravitational radius (assuming the pericenter distance of the star is comparable to the tidal radius of a solar-like star, being $\sim 0.5 R_{\odot}\left(M/M_{\odot}\right)^{1/3} \simeq 23 r_{\rm g}$ for a $10^6M_{\odot}$ black hole; \citealt{guillochon13, mainetti17, lawsmith20, nixon21}). Specifically, using Equation (3) from \citet{dexter19}, an $\alpha$-viscosity of 0.02, an accretion rate of $150 L_{\rm Edd}$, and a radius of $50 r_{\rm g}$ would yield an inflow time of $\sim 0.08$ days (which is why it is usually assumed that the early-time fallback in a TDE is tightly coupled to the accretion rate; \citealt{Rees1988, cannizzo90}). Requiring the inflow time to be $200$ days, on the other hand, and otherwise using the same values for the accretion rate and the radius would necessitate an $\alpha$-viscosity of $\alpha \simeq 8.1\times 10^{-6}$. This value of $\alpha$ is extremely small compared to what is inferred from magnetohydrodynamics simulations ($\alpha \simeq 0.01$; e.g., \citealt{hawley11}), and smaller yet compared to what is inferred from observations of dwarf novae ($\alpha \simeq few\times 0.1$; \citealt{king07}). 

However, simulations have shown that it can be difficult to remove the required (very large) amount of orbital energy to rapidly form a compact circular disk, and instead the debris may occupy an extended (typical radius of $\sim$1000 R$_g$), highly elliptical configuration (e.g. \citealt{Shiokawa}). Saturation of the magneto-rotational instability (MRI) happens on a timescale similar to that of a circular disk (e.g. \citealt{Chan2022}), and typically takes of order 10 orbital periods \citep{Shiokawa}. For typical parameters in an eccentric configuration this timescale is of order several 100 days, which is well matched with the X-ray delay seen here and in other systems. When accretion onto the SMBH is eventually triggered, the fall-back rate would be sufficiently high ($\sim$Eddington or higher) to explain the launch of a powerful outflow (see e.g. the discussion in \citealt{ryu2023}), as observed.

We conclude that without further constraints on the early-time X-ray data both scenarios remain plausible. 

\subsection{Radio emission (or a lack thereof) and outflows}
Radio observations could help to distinguish between the two scenarios described above. In particular, if the accretion rate is super-Eddington and the UFO is active at early times, then as the outflow/wind interacts with the circum-nuclear medium, it should produce detectable synchrotron emission if there is a sufficiently high circum-nuclear gas density and magnetic field.

Recent observations of a sizable sample of TDEs shows that while many sources are not promptly radio-bright, $\sim$ 50\% of these are detected in the radio with delays of 500-2000 days \citep{Cendes23}. This delay was interpreted as evidence for a lack of outflows around the UV/optical peak, for example due to a delay in accretion disk formation. The delay timescale points to outflow velocities of 0.02--0.15c when they do form 100s of days after the UV/optical peak.

It should be possible to apply a model for the supercritical accretion flow to assess the viability of the obscuration of the X-rays by a surrounding envelope, and the corresponding required observer line of sight. For example, the zero-Bernoulli accretion model \citep{coughlin14} proposes that the disc is quasi-spherical, with an angular density profile of $\left(\sin^2\theta\right)^{\alpha}$, where $\theta$ is the polar angle measured from the rotational axis and $\alpha$ (not to be confused with the Shakura-Sunyaev viscosity parameter) is a number that increases with the declining accretion rate (and thus the disc becomes less spherical and more disc-like with time; cf.~\citealt{wu18}). This model was recently employed by \citet{eyles22} in the context of jetted TDEs, though they only considered the purely on-axis and off-axis cases. The envelope should also provide a time-variable confinement for the outflow, where the confinement could either be due to shocks \citep{bromberg07, kohler12} or radiation-viscous interactions with the envelope \citep{coughlin20}, thus modifying the appearance of the outflow and the strength of the synchrotron emission. 

In the context of this picture, the observed delay in radio flares from TDEs is a feature that exposes the lack of dense circum-nuclear material in nuclei hosting TDEs. This is consistent with IR observations of TDE dust echoes, which point to very low dust covering fractions in TDE nuclei \citep{Jiang2021}. A typical TDE outflow with a velocity of $\sim$ 0.15c would reach distances from the SMBH of $\sim$0.1 pc in 1000 days. If a typical TDE clears out dust and gas on a similar scale (which is well matched to the typical dust sublimation radius of a TDE-like flare, \citealt{vanvelzen2016b}), one would not expect radio flares from shocks with the CNM until the outflow can reach $\sim$0.1 pc distances (unless the outflow itself produces jet-like radio emission, as was seen, for example, in ASASSN--14li; \citealt{vanvelzen2016}). 

One discrepancy between the observations and this interpretation is that the ZEBRA model proposes that the envelope is truncated at the trapping radius, implying that the optical/UV luminosity is capped at the Eddington limit of the black hole. This feature of the model is not consistent with the temporal evolution of the optical/UV emission of \target, which, as can be seen from Figure \ref{fig:lc}, declines by a factor of $\sim 5$ across all bands over $\sim 400$ days; the same inconsistency was noted by \citet{eyles22}. This discrepancy could imply that the photosphere is at a larger distance from the black hole than the trapping radius (this would also reconcile the fact that ZEBRA temperatures are a factor of a few larger than those inferred from observations; \citealt{eyles22}), that the time-dependent evolution of the disc geometry (coupled to the observer line of sight) alters the observed optical/UV luminosity in a way that is decoupled from the emitted luminosity (which is presumed to be capped at the Eddington limit), or the amount of energy processed and re-radiated by the envelope declines with the fallback rate. More detailed and numerical modeling could address which one of these could alleviate this tension.
\begin{figure*}[h]
    \centering
    \includegraphics[width=0.45\textwidth]{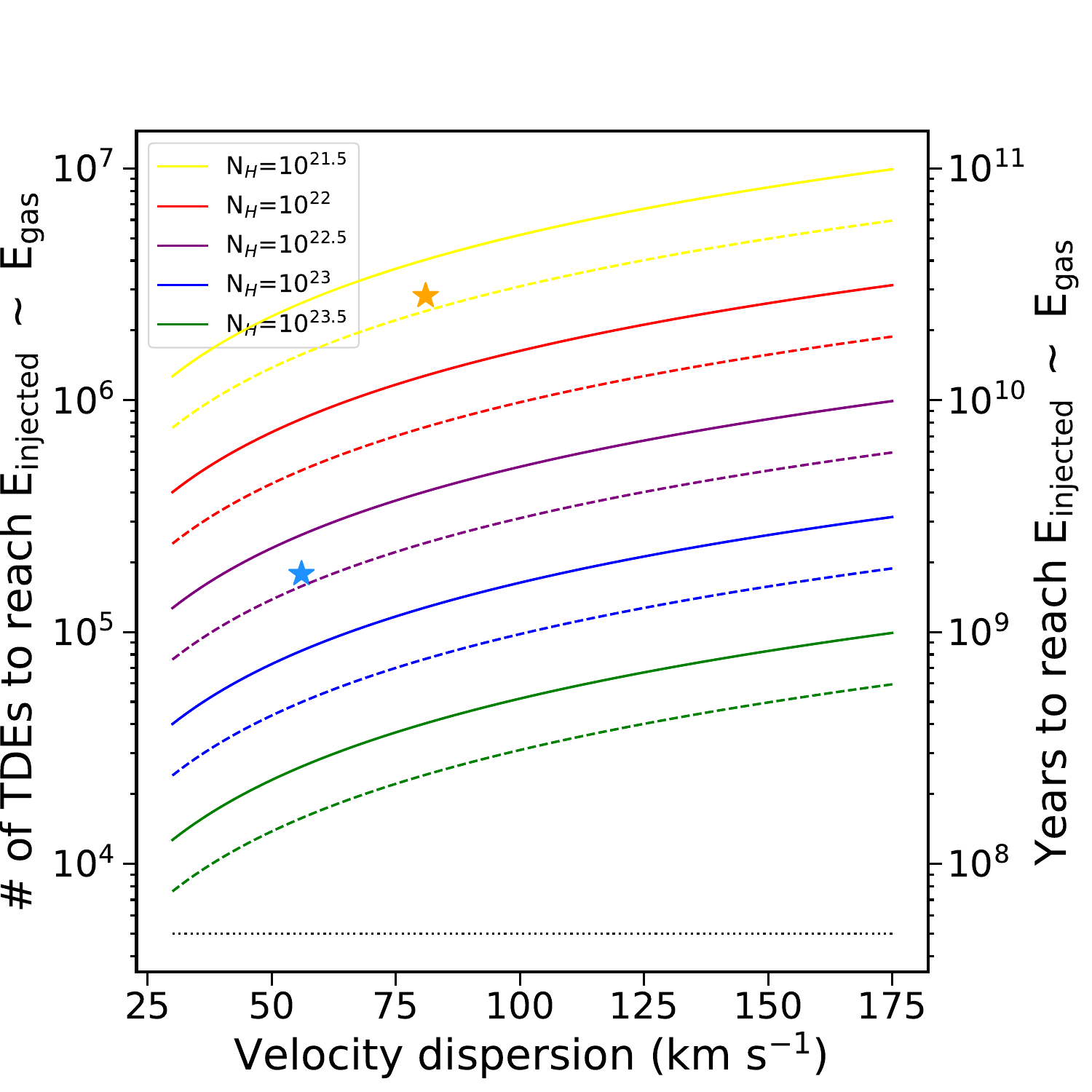}
    \includegraphics[width=0.45\textwidth]{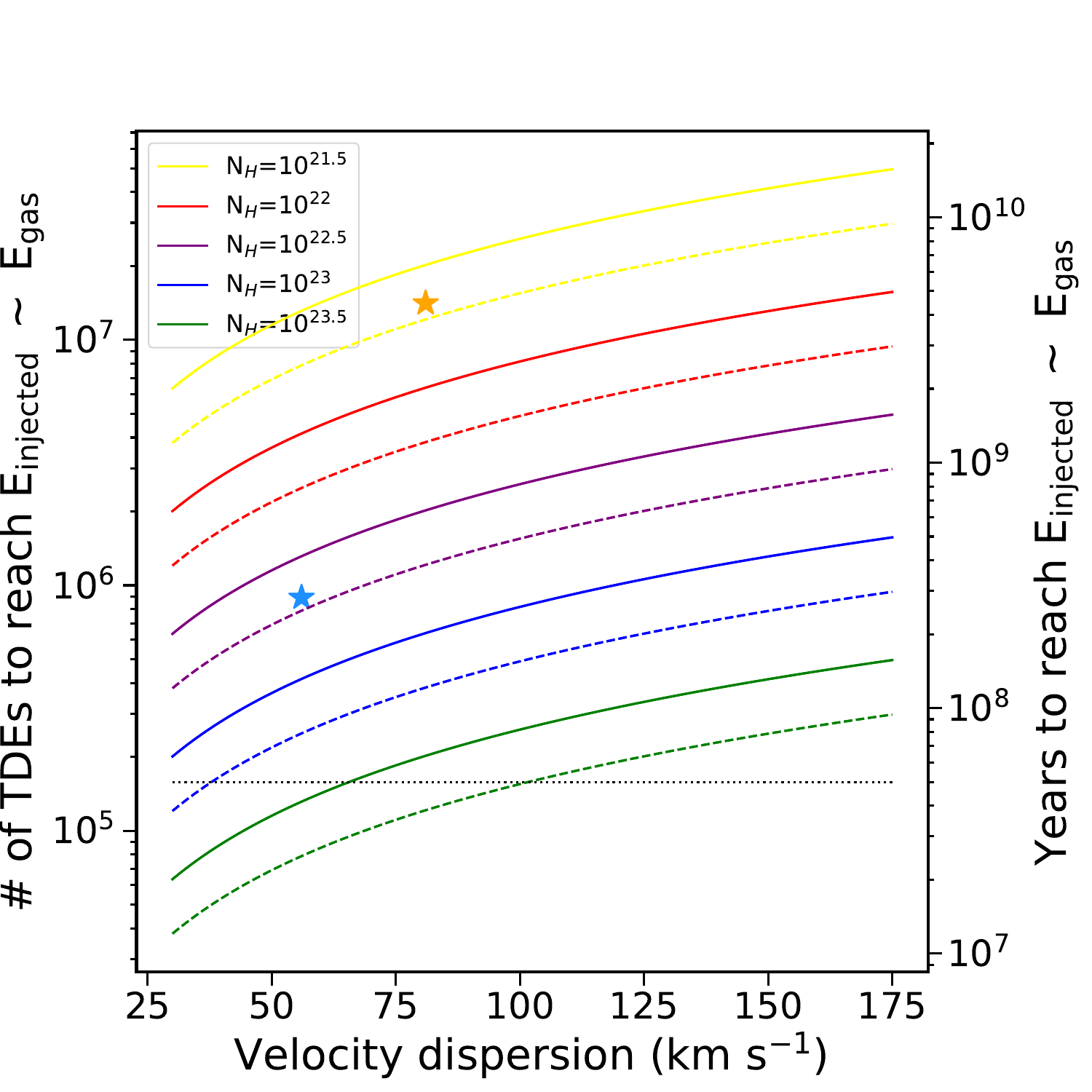}
    \caption{{\bf Energy budget of TDE wind feedback.} {\it Left panel:} the number of TDEs required to inject an energy comparable to the gas mass gravitational binding energy as a function of velocity dispersion (assuming a gas mass fraction f$_{\rm gas}$ = 0.1). The alternative vertical axis converts this to years by assuming a TDE rate of 10$^{-4}$ per galaxy per year. Solid (dashed) lines indicate v$_{\rm wind} = 0.15$c (0.25c), for various wind column densities. The blue and orange stars represent estimates for AT2020ksf and ASASSN--14li-like UFOs. The dashed horizontal line represents the accretion Salpeter time. {\it Right panel:} Same, but for a higher gas mass fraction and TDE rate, more typical of post-starburst (E+A) galaxies. }
    \label{fig:feedback}
\end{figure*}
\subsection{TDE feedback to the host galaxy}
\label{sec:feedback}

Recalling that P$_{\rm kin} \propto \rm N_H v^{3}_{w}$, when accounting for the difference in velocity and column density between the two systems, we infer that P$_{\rm kin}$ is a factor of $\sim$10 higher in AT2020ksf than in ASASSN--14li (Ajay et al. submitted). 

For a more quantitative estimate, we consider the number of TDEs required to inject an energy comparable to the gravitational binding energy of the stellar bulge (E$_{\rm bulge}$). We make several generalizing assumptions, the most important ones being a UFO duration of 1 year and a UFO fraction of 1 (i.e. each TDE launches a UFO); we further assume a typical wind velocity of 0.15c (as observed for both AT2020ksf and ASASSN--14li). We find that the wind energy budget is orders of magnitude smaller than E$_{\rm bulge}$; it would require $\sim$ 10$^{5-7}$ TDEs to inject a comparable amount of energy. This would take longer than a Hubble time assuming a TDE rate of 10$^{-4}$ per galaxy per year. We note that this number could be substantially higher if the early X-ray dark period is related to a higher column density wind obscuring the X-ray emission in the first 230 days, as discussed in the previous section. 

While the wind energy budget is too small to influence the stellar bulge, many galaxy nuclei are also home to neutral and ionized gas in their bulges. To compare the wind energy to the gas binding energy, we assume a gas mass fraction of f$_{\rm gas}$ = 0.1 (typical for Milky Way like galaxies and galaxy stellar masses of 10$^{10}$ M$_{\odot}$, e.g. \citealt{Ellison2018}). This is illustrated in the left panel of Fig. \ref{fig:feedback}, where the solid (dashed) lines indicate a wind velocity of 0.15c (0.25c) for varying wind column densities. The best-fit values for AT2020ksf and ASASSN--14li are shown as blue and orange stars, respectively. Although the timescales are more favorable in this scenario, they remain longer than a Gyr, which is likely too long for TDEs to significantly influence the nuclear gas reservoir.

Finally we consider the special case of E+A galaxies, which are known to be over-represented among TDE host galaxies by 1--2 orders of magnitude (e.g. \citealt{French2020}). Taking into account the likely elevated gas fraction (f$_{\rm gas}$ = 0.5) and elevated TDE rate of 10$^{-2.5}$ per galaxy per year \citep{Stone2016, French2020} we find that in this case the timescales decrease significantly, and the energy budgets can be matched if the elevated TDE rate can be sustained for a few $\times$10$^8$ years (Fig. \ref{fig:feedback} right panel).
Telltale signs of this effect could include increased nuclear gas temperatures/turbulence and/or large ($\sim$ kpc-scale) but much slower (100s of km s$^{-1}$) outflows, observable through optical/radio observations. If confirmed, this might help to provide an explanation for the rapid depletion of the molecular gas content in the post-starburst evolution of E+A/post-merger galaxies \citep{French18}.

Several important notes are in order when interpreting these calculations. First, in the absence of meaningful constraints we have optimistically assumed that every TDE will launch a UFO. Second, we have assumed that the feedback introduced by the TDE (whether directly by the ram pressure of the wind, or through radiative and/or inverse Compton cooling) is 100\% efficient. A less efficient coupling would decrease the available energy budget; fiducial values are coupling efficiencies of 5\%, although with large scatter, for AGN wind feedback (e.g. \citealt{King2015, Harrison2018}). 
These considerations could further lower the energy injection, hence decreasing the potential of TDEs to contribute to galaxy feedback.

On the other hand, there are several situations that would help to increase the (integrated) energy budget. First, X-ray observations are sensitive only to the column density along our line of sight. Any direct comparison between different sources should be done by keeping in mind that different orientation angles may probe different parts of the wind. For example, if for ASASSN--14li we have a close to pole-on view, our inferences would likely be lower limits to the true wind densities and hence kinetic power, as the density is expected to increase towards the disk plane.
Second, if the X-ray delay is due to additional ionized absorption at early times, the wind kinetic power estimates for AT2020ksf could increase by more than an order of magnitude because the kinetic energy injection scales linearly with the integrated wind column density. If the wind column density in AT2020ksf was $>$10$^{23.9}$ cm$^{-2}$ (as required to explain the X-ray non-detection with an ionized absorber), the energy budget for feedback would be $\sim$20 times higher.Systematic follow-up observations of TDEs will reveal whether some TDEs are able to launch more powerful UFOs than AT2020ksf, which would be able to inject more energy into the circum-nuclear gas.

Third, if a UFO is present for a significantly longer timescale, then the available energy would rise proportionally. Current X-ray instruments are not sensitive enough to probe faint absorption features at the low flux levels of observed TDEs at late times. Deep, late-time monitoring is required to study the long-term evolution of UFOs in TDEs.

\section{Summary}
\label{sec:summary}
We have reported on the discovery of transient absorption features in the X-ray spectra of the tidal disruption event Gaia20cjk/AT2020ksf. The TDE is not X-ray bright at UV/optical peak, but brightens by at least a factor of 25 in X-rays (compared to the UV/optical peak upper limit) 230 days later; with an inferred black hole mass $\sim$10$^6$ M$_{\odot}$, the X-ray emission is at the Eddington limit when the absorption features are first present. An absorption feature remains detectable up to 770 days after peak. Based on AIC analysis of a variety of models, including combinations of simple continuum models, a partial covering model and a high velocity ionized absorber model, we find that the fast ionized absorber provides the best description when considering the early and late time data together; a partial covering model can describe only one of the two instances well. We therefore propose a physical interpretation as a highly ionized outflow with a velocity up to --0.15c, whose column density and ionization parameter decrease over time. This is only the second UFO to be reported in a TDE; the discovery is facilitated by the excellent soft X-ray sensitivity of the NICER XTI instrument. By binning high temporal coverage data spanning 7 days in 2 epochs, we find that in the first epoch an ionized high velocity absorber improves the fit but is not statistically preferred. During the second epoch several days later, the absorber is clearly present. The absence of an outflow signature during the first epoch implies rapid variability of the disk wind on a timescale of a few days. 

We discussed two scenarios to explain the properties of both AT2020ksf and its twin TDE ASASSN--14li, including a difference in the UFO properties, or a difference in disk formation efficiency. The presence of a 200 day gap in the X-ray data precludes us from distinguishing between these scenarios. 

In terms of energy injection / wind feedback on the host galaxy gas, the two known TDE UFOs differ by more than an order of magnitude; further exploration is required to fully establish the energy budgets of TDEs through detailed UFO modeling of future X-ray bright sources. Through quantitative estimates of the TDE wind feedback, we find that TDEs may play a significant role in galaxy feedback if the TDE rates can be elevated for an extended period of time, for example during the post-starburst evolution of E+A/post-merger galaxies. 

Late-time data taken 770 days after UV/optical peak also show evidence for an outflow, similar to that observed in ASASSN--14li at peak. This second serendipitous detection of a UFO, displaying rapid variability on several day timescales, suggests that systematic X-ray follow-up observations at multiple phases with respect to the UV/optical peak may uncover that a significant number of TDEs are launching powerful outflows. We further highlight that AT2020ksf is a factor $\approx 5$ more distant than ASASSN--14li, but located near the median redshift of the homogeneously selected ZTF sample \citep{Hammerstein2023}. To assess whether the lack of UFO detections in the literature is related to the poor sensitivity and sampling of existing X-ray follow-up observations, or has a physical origin, requires a systematic follow-up survey. This will have important implications for the ability of accretion flows to launch outflows, and the potential production of neutrino emission in TDEs (e.g. \citealt{murase, guepin}; see also \citealt{Wevers23} and references there-in).

\begin{acknowledgments}
We thank the anonymous referee for comments and suggestions that improved the paper.
TW is grateful to the Space Telescope Science Institute, where part of this work was completed during a research visit, for its hospitality. TW thanks Y. Sun, A. Zabludoff, N. Stone and J. Krolik for insightful discussions. We thank the \swift\ and \nicer\ PIs and their operations teams for approving and promptly scheduling the requested observations. Raw optical/UV/X-ray observations are available in the NASA/\swift\ archive (\url{http://heasarc.nasa.gov/docs/swift/archive}, Target Names: AT2020ksf). \nicer\ data is publicly available through the HEASARC: \url{https://heasarc.gsfc.nasa.gov/cgi-bin/W3Browse/w3browse.pl}. 
ERC acknowledges support from the National Science Foundation through grant AST-2006684, and from the Oakridge Associated Universities through a Ralph E. Powe Junior Faculty Enhancement Award. The authors wish to recognize and acknowledge the very significant cultural role and reverence that the summit of Mauna Kea has always had within the indigenous Hawaiian community. We are most fortunate to have the opportunity to conduct observations from this mountain.
This work made use of data supplied by the UK Swift Science Data Centre at the University of Leicester.
\end{acknowledgments}

\vspace{5mm}
\facilities{NICER, Swift(XRT and UVOT), XMM-Newton, eROSITA, Keck}

\FloatBarrier

\newpage
\appendix
\setcounter{table}{0}
\renewcommand{\thetable}{A\arabic{table}}
\setcounter{figure}{0}
\renewcommand{\thefigure}{A\arabic{figure}}

\section*{Supplementary material}
\section{Data Reduction and Analysis}
\label{sec:data}
\begin{figure}
    \centering
    \includegraphics[width=0.32\textwidth]{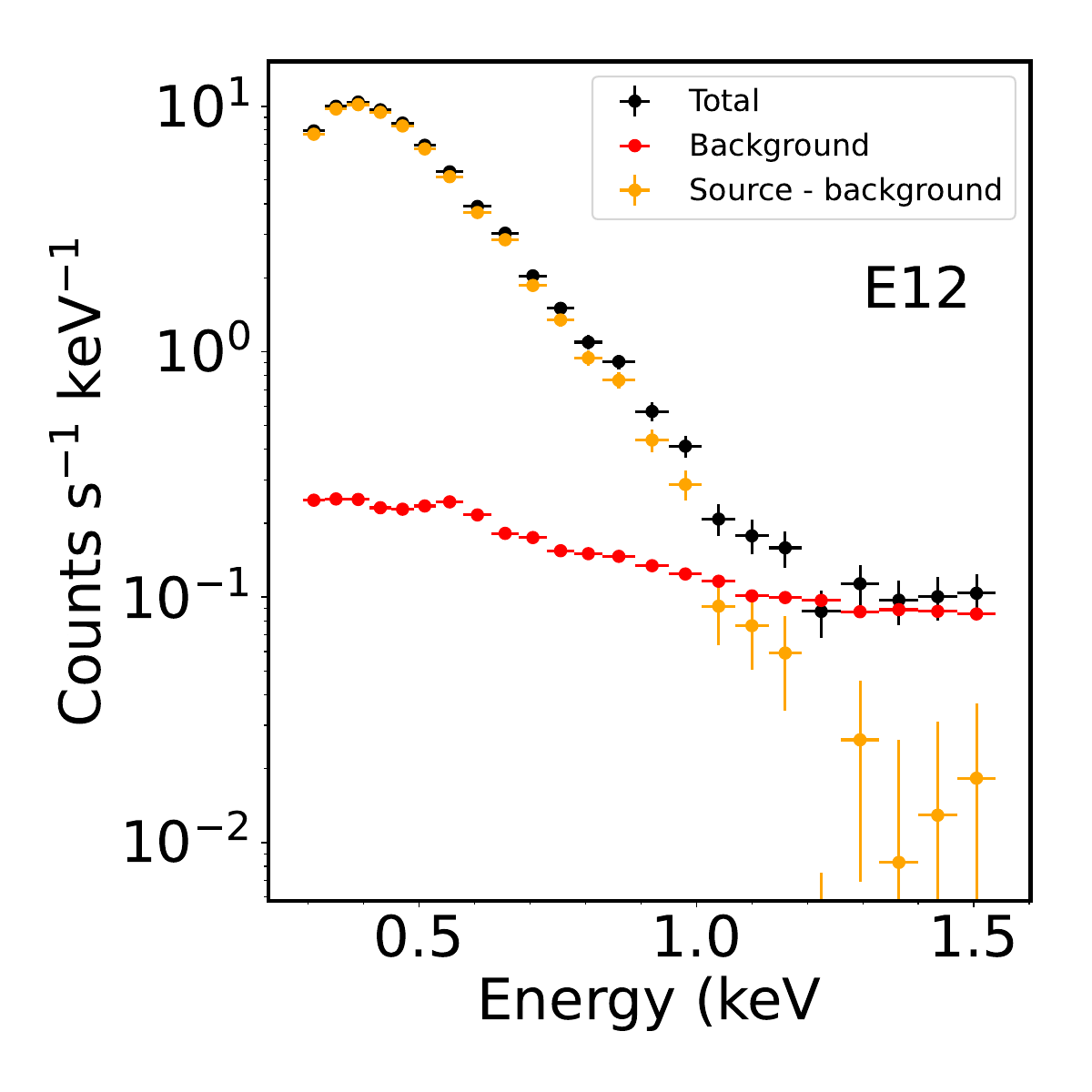}
    \includegraphics[width=0.32\textwidth]{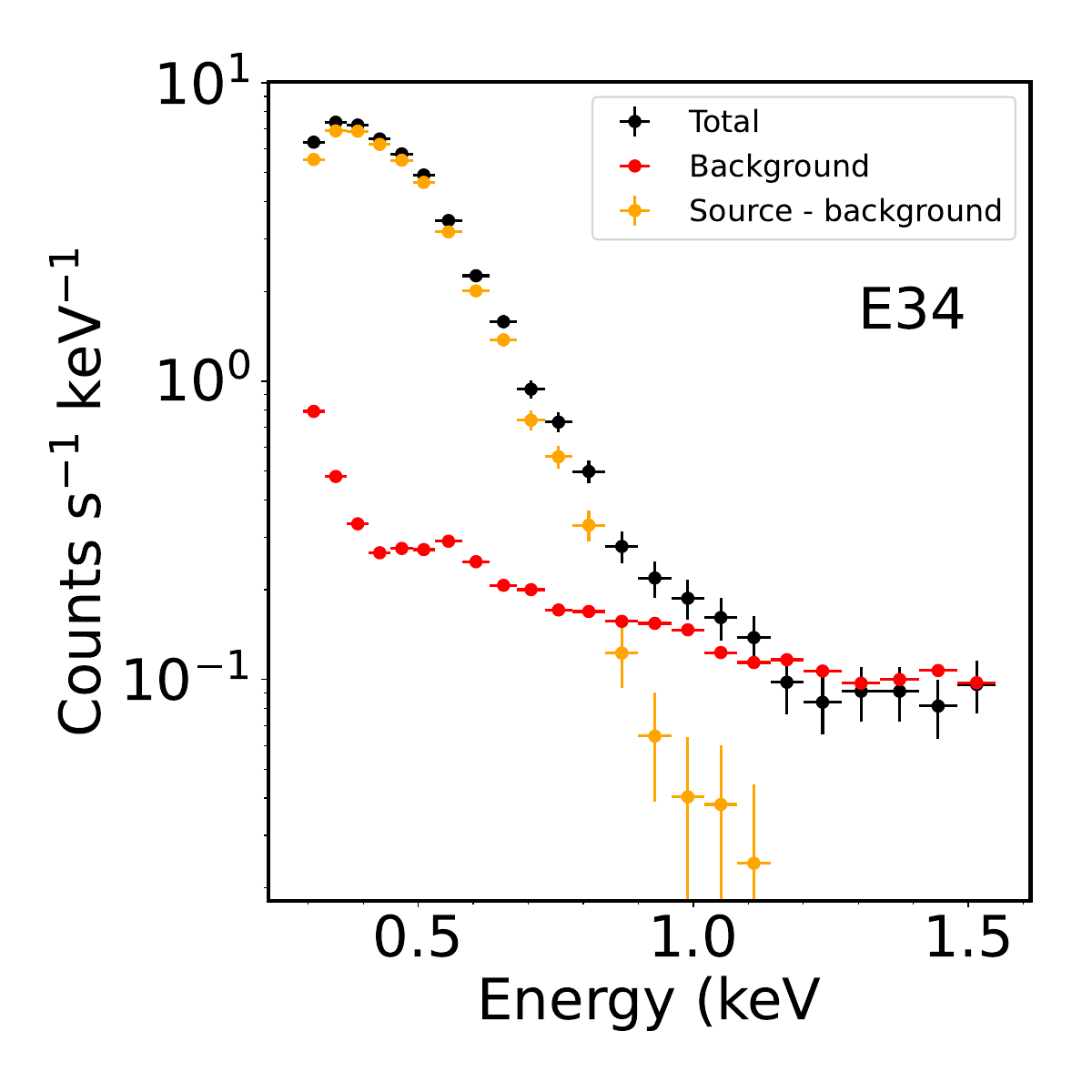}
    \includegraphics[width=0.32\textwidth]{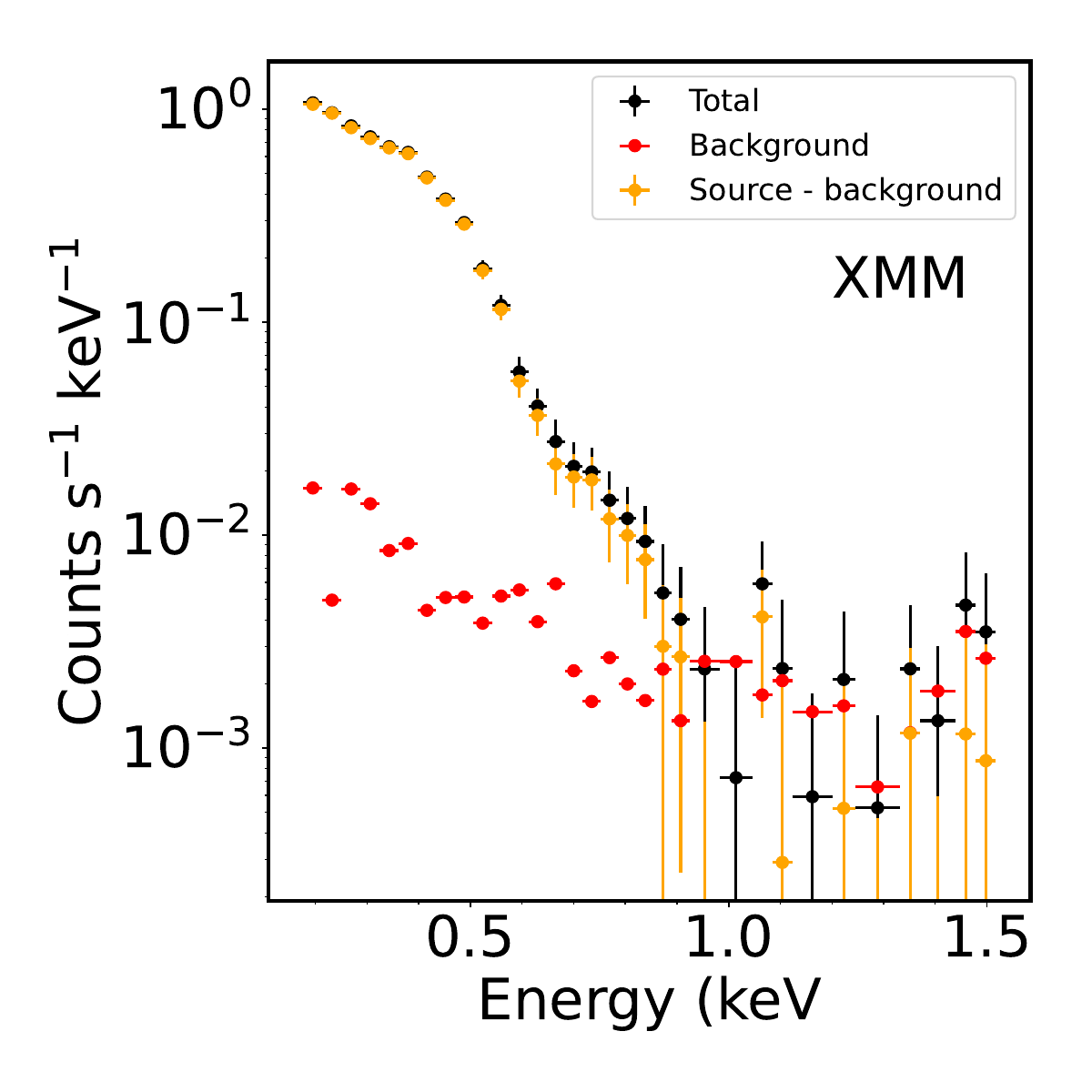}
    \caption{Source and background count rates for the NICER E12 (left) and E34 (right) and XMM-Newton (right) epochs. We use the energy range between 0.3--1.1 keV for fitting the NICER spectra, and 0.2--0.9 keV for XMM-Newton. We carefully checked that our results are not influenced by the exact energy range used. }
    \label{fig:background}
\end{figure}

We describe the X-ray data reduction in detail in this Section; the resulting spectra are shown in Figure \ref{fig:background}.

\subsection{\nicer}\label{sec:nicer}
Following the X-ray detection by eROSITA \citep{2020ATel14246....1G} \nicer\ started observing \target\ on 2020 December 4 (MJD 59187) as part of an approved guest observer program (PI: Pasham). It took multiple exposures of duration between 300 and 1400 seconds for 8 days before the source became sun constrained, totalling 10 ks. \target\ was observed again on 2022 April 30, and a third set of exposures were taken more recently on 2022 October 28--29. The total \nicer\ exposure time for AT2020ksf is 20 ks.

We started our \nicer\ data analysis by downloading the raw, so-called level-1, data available on the High Energy Astrophysics Science Archive Research Center (HEASARC). Data reduction was performed with the standard {\it NICER} Data Analysis Software (NICERDAS/HEASoft 6.29c). The data were reduced using the {\it nicerl2} task with the standard steps outlined in the \nicer\ data analysis guide: \href{https://heasarc.gsfc.nasa.gov/docs/nicer/analysis_threads/nicerl2/}{https://heasarc.gsfc.nasa.gov/docs/nicer/analysis\_threads/nicerl2/}. Good Time Intervals (GTIs) were extracted with default values for all parameters except {\it underonly\_range}, {\it overonly\_range}, {\it overonly\_expr} which were set to accept all values. Background spectra were estimated on a per-GTI basis using the 3c50 model \citep{3c50}. The individual GTIs were marked as acceptable only if the background-subtracted values of the so-called S0 band (0.2-0.3 keV) and HBG band (13-15 keV) were above 2 counts s$^{-1}$ and 0.05 counts s$^{-1}$, respectively. This is referred to as the level-2 filtering as per \citet{3c50}. More details of this analysis procedure can be found in \citet{2022cmc}.

In recent observations there is a transient telluric oxygen line (the strength of which depends on the viewing angle of the source with respect to the Earth's magnetic poles among other unknown factors) that can potentially contaminate the energy region of interest here ($\sim$0.4--1 keV). 

\subsubsection{Monitoring the potentially contaminating Oxygen line from Earth's atmosphere using PSR~B1937+21 with NICER}
\begin{figure}
    \centering
    \includegraphics[width=0.7\textwidth]{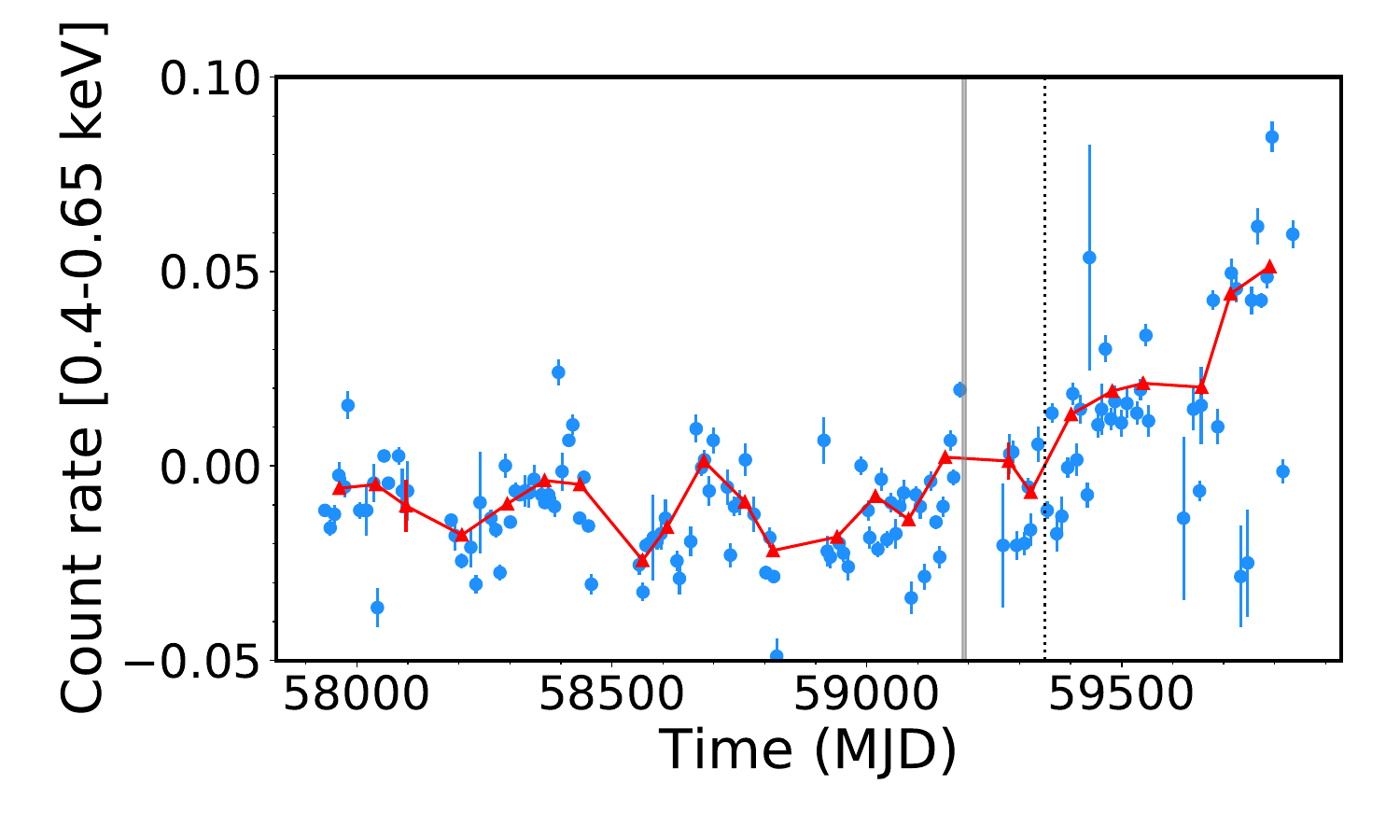}
    \caption{{\bf Long-term evolution of the foreground Oxygen line}. The y-axis shows the background-subtracted 0.45-0.6 keV count rate of the pulsar PSR~B1937+21, which is one of the prime \nicer targets and has been observed throughout the mission operations. Each point represents the mean value within a Good Time Interval (GTI) . Systematic deviations above the background beyond 1$\sigma$ are evident past MJD 59350 (dashed line, May 2021), which is well after the observations of AT2020ksf presented in this work (grey band). }
    \label{fig:psrmonitor}
\end{figure}

As part of its roughly 11 year cycle, our Sun's activity has been increasing. One consequence of this is the puffing up of the Earth's atmosphere, which can occasionally intersect \nicer's line of sight to an astrophysical target. This can manifest as a broad, telluric Oxygen emission line with zero width whose centroid is constrained to be between 0.5--0.6 keV. If the astrophysical target also has an Oxygen feature around the same energy then it becomes challenging to model the spectrum. At present, the \nicer team does not have a tool to predict the times when this happens. However, we can track the impact of this foreground contamination over the lifetime of the mission using known persistent/stable targets and show that when \nicer observed \target this contamination was not pronounced in \nicer datasets. In Figure \ref{fig:psrmonitor}, we show 3c50 background-subtracted 0.45-0.6 keV lightcurves of PSR~B1937+21, a very stable pulsar that has been observed routinely by NICER since the beginning of science operations in 2017. If there were no contamination this lightcurve should be consistent with 0. But it is evident that the count rate started rising after $\approx$ MJD 59350 (marked with a dashed black line). \nicer observations of \target were taken much earlier than that epoch when PSR~B1937+21's background-subtracted 0.4-0.65 keV light curve is consistent with the background, i.e., no evidence for foreground contamination. 

This establishes that the absorption feature seen in NICER data (marked by the grey band) did not originate from Earth's atmosphere.

\subsubsection{Investigating another TDE spectrum taken with NICER in the E34 time window}
\begin{figure}
    \centering
    \includegraphics[width=0.7\textwidth]{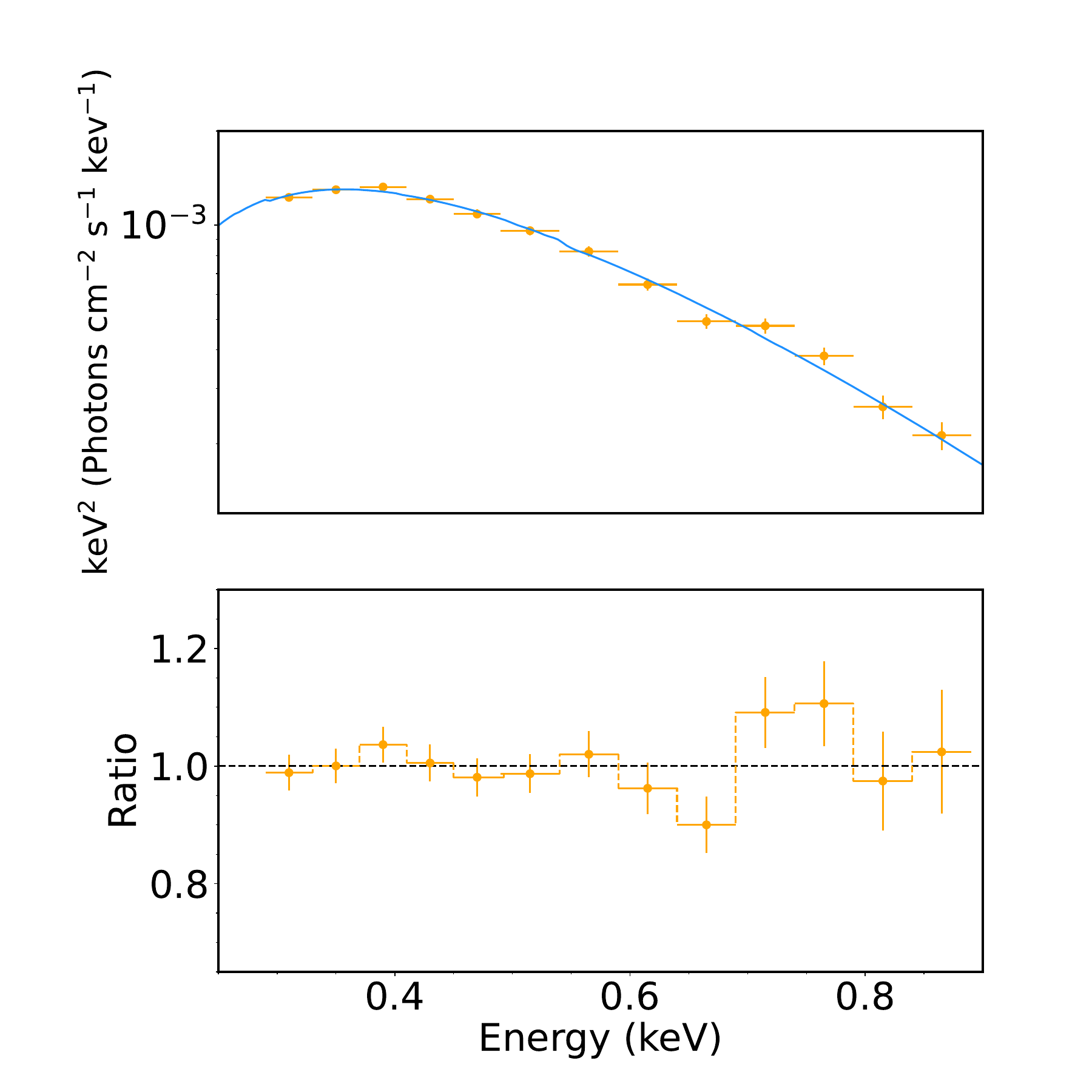}
    \caption{Stacked spectrum of AT2020ocn (top), and ratio of a thermal model fit to the continuum (bottom). }
    \label{fig:ocn}
\end{figure}

To further investigate this feature, we also reduced data of another TDE, AT2020ocn. NICER data of this source overlapping with the E34 epoch of AT2020ksf was stacked. The source has a very soft spectrum and contains a similar number of X-ray photons ($\approx$ 12\,000). A thermal continuum fit to the energy range 0.3--0.8 keV (where the source is detected above the background) results in a fit statistic of 12 for 10 degrees of freedom. Adding a Gaussian absorption line ({\tt gabs} in Xspec) to assess the presence of an absorption feature, we find that the improvement in fit statistic is 5 for 3 degrees of freedom (line energy, width and strength). This is equivalent to a $\Delta$(AIC) = +1, i.e. the absorption feature is not statistically significant. The spectrum, model fit and ratio are shown in Figure \ref{fig:ocn}. Combined with the presence of a similar absorption feature seen in the XMM-Newton data indicates that the absorption feature in the AT2020ksf E34 data is real.

\subsubsection{Absorption line model and $\chi^2$ contour maps}
\label{sec:chi2map}
\begin{figure}
    \centering
    \includegraphics[width=\textwidth]{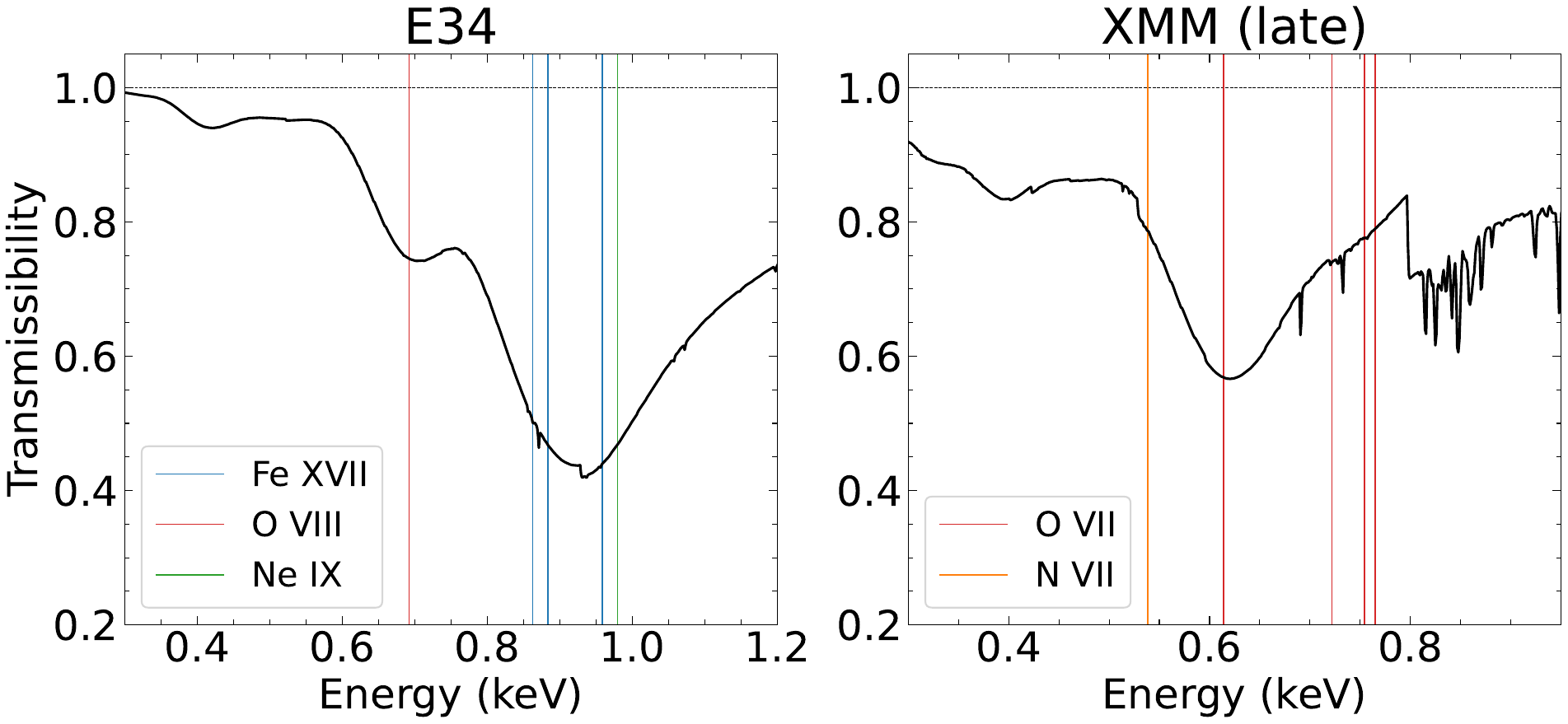}
    \caption{Absorption line profiles corresponding to the best-fit UFO parameters for E34 (left) and XMM (right), with the identification of the 5 strongest transitions that are causing the absorption features in the observed spectrum. Note that the horizontal axis denotes the observer frame. The narrow features (most prominently visible in the right panel) are due to absorption edges which are not broadened by the {\tt xstar} code, and do not influence the fit because they are very narrow and relatively weak.}
    \label{fig:ionline}
\end{figure}

We show the ratio between the data and the model (including the absorber) for epoch E34 and XMM in Figure \ref{fig:ratio}. 
To identify the ionic species responsible for the (blueshifted) absorption feature, we plot the absorption line profile corresponding to the best-fit UFO parameters in Figure \ref{fig:ionline}.

\begin{figure}
    \centering
    \includegraphics[width=0.45\textwidth]{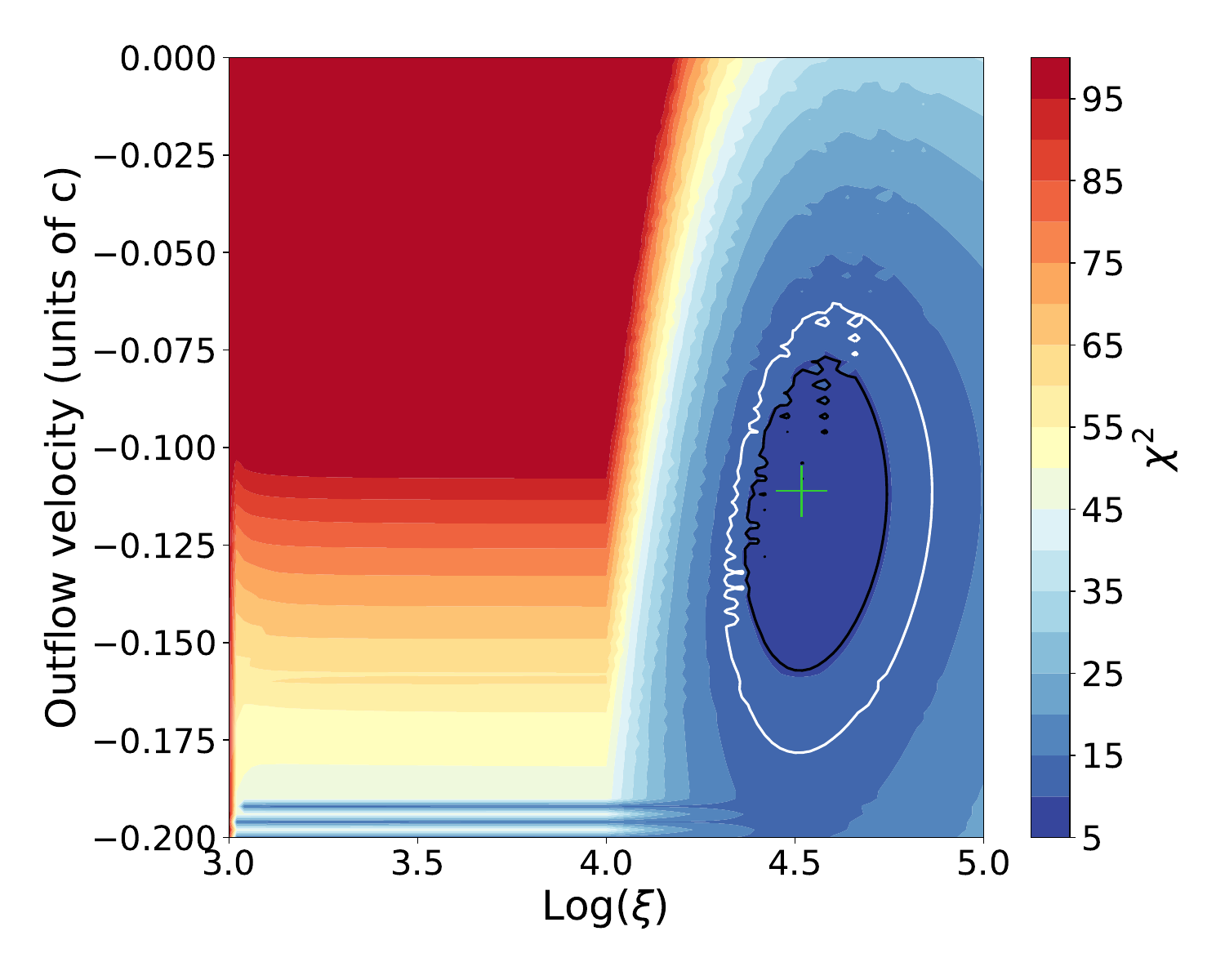}
    \includegraphics[width=0.45\textwidth]{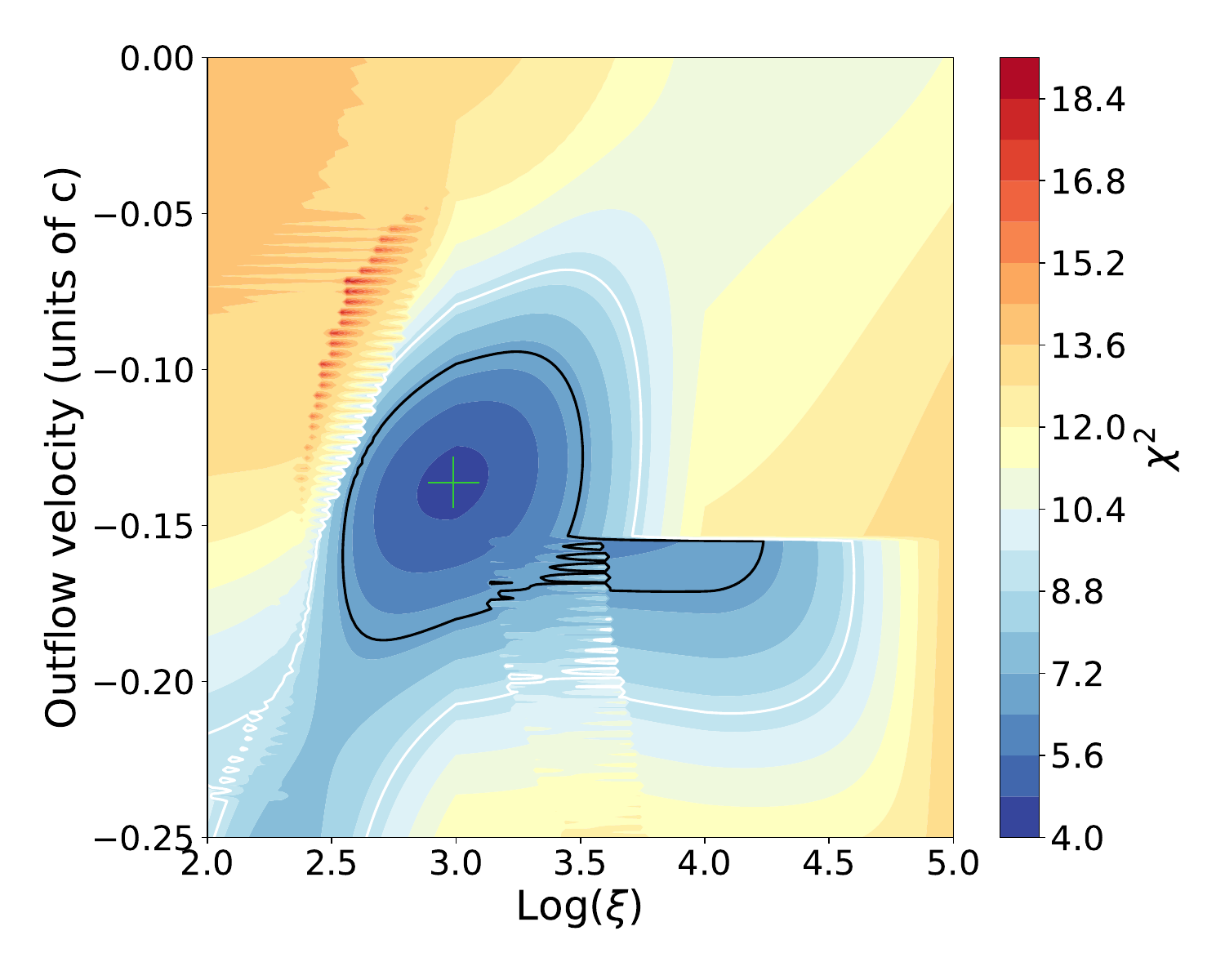}
    \includegraphics[width=0.45\textwidth]{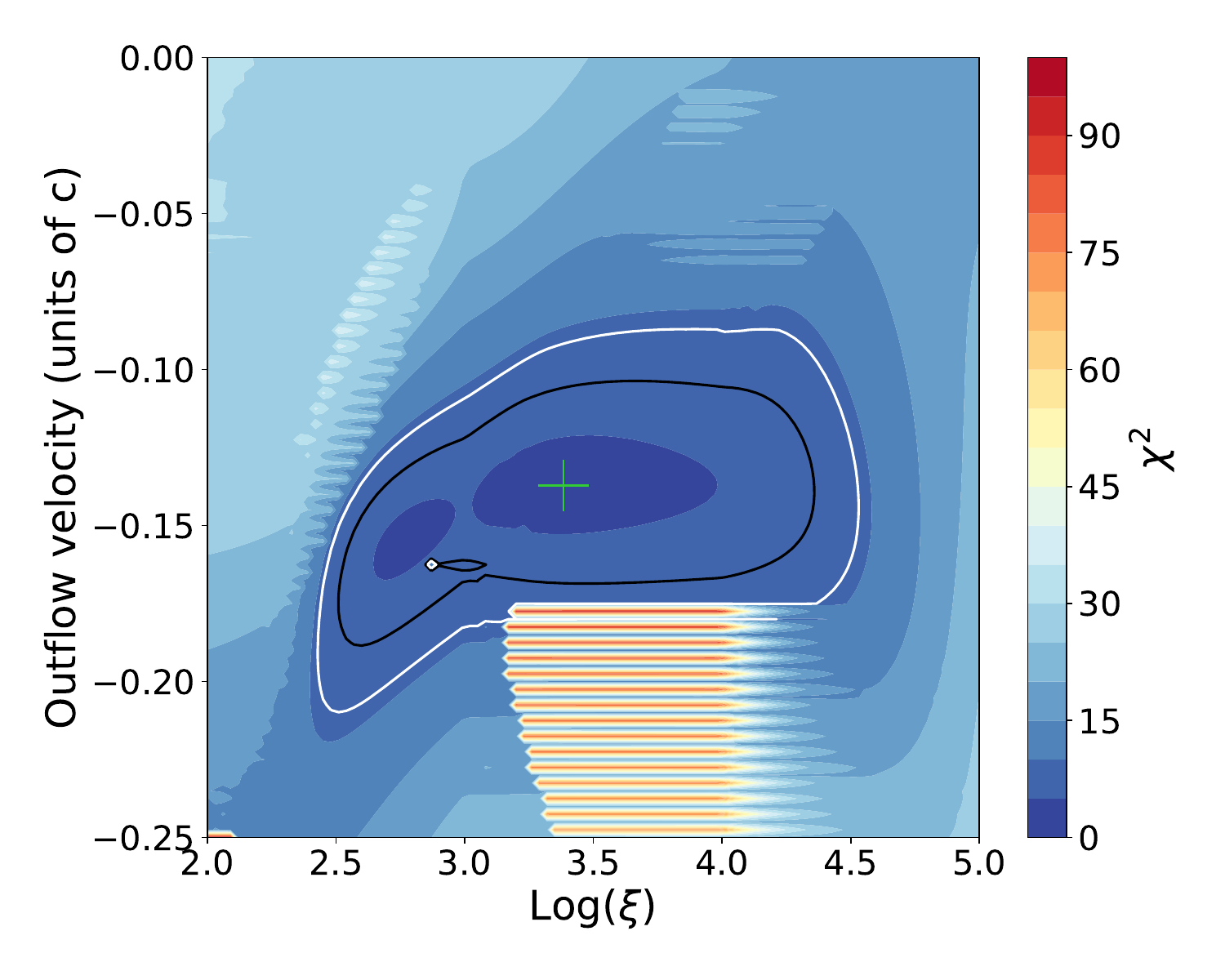}
    \caption{2D fit statistic contour plot for the ionized outflow parameters. The green cross marks the best-fit value, and the solid black and white lines indicate the 68\% and 90\% confidence contours. The top two panels show the individual epoch spectra (E3 on the left and E4 on the right), while the bottom panel shows the combined spectrum (E34)}
    \label{fig:contours}
\end{figure}

We generate the 2D map with the $\Delta \chi^2$ contours for epoch E34 in which we significantly detect the UFO, to allow an assessment of the typical parameter degeneracy. These are shown in Figure \ref{fig:contours}. For the combined spectrum, the $\chi^2$ landscape looks well-behaved with a clear global minimum. 

\subsection{\swift/XRT}\label{sec:swift}
\swift\ observed \target\ on $\sim$20 occasions between MJD 59180 and MJD 59900. These observations were reduced using the standard {\it xrtpipeline} task as recommended on \swift's data analysis page: \href{https://www.swift.ac.uk/analysis/xrt/}{https://www.swift.ac.uk/analysis/xrt/}. The cleaned eventfiles from the the above step were used for further analysis with the additional filter to only use events with grades between 0 and 12. Source events were extracted using an circular extraction region with a radius depending on the uncorrected count rate \citep{Evans2009}, centered on optical coordinates (RA, Dec) J2000.0 = (323.863583, --18.276539). The background was estimated using a annular region centered on the above position with an inner and outer radii of 142\arcsec and 260\arcsec, respectively.  
To convert from background-subtracted count rate to luminosity we extracted an average X-ray spectrum by combining all the existing XRT data. We then fit it with a thermal (disk blackbody) model ({\it tbabs*zashift(diskbb)} in {\tt Xspec}, \citealt{xspec}). The mean count rate and observed 0.3-1.1 keV flux were 0.022 cps and 1.67$\times$10$^{-12}$ erg~s$^{-1}$, respectively. From this we derived a scaling factor of 9.44$\times$10$^{-12}$. This provides consistency in the X-ray luminosity for the epochs with contemporaneous XRT and XTI data around 200 days after peak (Figure \ref{fig:lc}).

\subsection{\xmm/EPIC}\label{sec:xmm}
\xmm\ observed \target\ on 2022 May 26 (MJD 59725), i.e., 770 days after optical discovery, for a duration of 31 ks. This data is part of an approved GO program (PI: Gezari, obsID 0882591201). The observation data files (ODFs) were reduced using the \xmm\ Standard Analysis Software \citep[SAS;][]{Gabriel_04}.
The raw data files were then processed using the \texttt{epproc} task. 
Since the pn instrument has larger effective area than MOS1 and MOS2, we only analyze the pn data. 
Following the \xmm\ data analysis guide, to check for background activity and generate ``good time intervals'' (GTIs), we manually inspected the background light curves in the 10--12\,keV band. We reject a 5 ks time interval where the background count rate indicated a large flare, and furthermore select only instrumental GTIs.
Using the \texttt{evselect} task, we only retained patterns that correspond to single and double events (\texttt{PATTERN$<$=4}). The source spectra were extracted using a source region of $r_{\rm src} = 35^{\prime\prime}$ around the peak of the emission. 
The background spectra were extracted from a $r_{\rm bkg} =
100^{\prime\prime}$ region located in the same CCD. The ARFs and RMF files
were created using the \texttt{arfgen} and \texttt{rmfgen} tasks,
respectively. 

We note that two fortuitously timed X-ray observations, by the XMM-Newton slew survey and eROSITA \citep{2020ATel14246....1G}, were taken contemporaneously with the UV/optical peak of emission, on MJD 58972 and 58978 (phase +12 and +18 days with respect to the discovery date). The most constraining upper limit is provided by eROSITA, where the converted 0.3--1.1 keV luminosities (using webPIMMS) are constrained to be lower than 8$\times$10$^{42}$ erg s$^{-1}$ (at 3$\sigma$, assuming a 100 eV disk blackbody spectral model and Galactic n$_{\rm H}$ \citep{2020ATel14246....1G}). Assuming instead a 70 eV single-temperature blackbody, or a 100 eV broader thermal ({\tt brehmsstrahlung}) model increases this upper limit by $\sim$15\%.  We can therefore rule out that there was luminous X-ray emission around the UV/optical peak, implying that the source brightened by a factor of $\gtrsim$25 on a timescale of $\sim$200 days. An archival ROSAT upper limit is also available, constraining the prior X-ray emission to $<$10$^{43}$ erg s$^{-1}$ (3$\sigma$ in the 0.3--1.1 keV band); another five XMM-Newton slew observations taken between these two dates provide similar upper limits on any X-ray emission prior to the UV/optical flare.

\subsection{UV and optical photometry}
UV observations were taken with \swift/UVOT contemporaneously with the XRT observations. 
We used the \texttt{uvotsource} package to measure the UV photometry, using an aperture of 5\arcsec. We subtracted the host galaxy contribution by modeling archival photometry data with stellar population synthesis using \textsc{Prospector} \citep{Johnson_21}, following the procedure described in \citep{Wevers2022} and tabulated in Table \ref{tab:hostgalphot}. We apply Galactic extinction correction to all bands using $E(B-V)$ values from \citet{Schlafly2011}. We corrected all photometry for the host galaxy contribution and Galactic reddening of E(B--V) = 0.04.

\begin{table}
    \centering
    \begin{tabular}{c|cc}
        Band & Observed & Model\\
        & (AB mag) & (AB mag) \\\hline
       PS1 $g$ & 18.63 (0.01) & 18.62	0.01) \\
       PS1 $r$ & 17.98 (0.01)& 18.01 (0.01) \\
       PS1 $i$ & 17.69 (0.01)& 17.70 (0.01) \\
       PS1 $z$ & 17.52 (0.02)& 17.50 (0.01) \\
       PS1 $y$ & 17.54 (0.04)& 17.41 (0.01) \\
       2MASS $J$ & 17.38 (0.25)& 17.25 (0.01) \\
       2MASS $H$ &16.88 (0.23) & 17.11 (0.02) \\
       2MASS $K_{\rm s}$ & 16.9 (0.4)& 17.28 (0.02) \\
       WISE $W1$ & 17.80 (0.04)&  17.79 (0.04) \\
       WISE $W2$ & 18.21 (0.07)& 18.39 (0.04\\\hline
       UVOT $U$ & ---& 20.47 (0.09) \\
       UVOT $B$ & ---& 19.11 (0.02)  \\
       UVOT $V$ & ---& 18.29 (0.01)  \\
       UVOT $UVW2$ & ---& 23.10 (0.99) \\
       UVOT $UVM2$ & ---& 22.73 (0.70) \\
       UVOT $UVW1$ & ---& 21.90 (0.31) \\
    \end{tabular}
    \caption{Results of the host SED model fitting. Values between brackets indicate the uncertainties, which are propagated into the host subtracted photometry.}
    \label{tab:hostgalphot}
\end{table}

We performed point spread function (PSF) photometry on all publicly available Zwicky Transient Facility (ZTF) data using the ZTF forced-photometry service \citep{Masci2019} in $g$- and $r$-bands, and $o$-band from ATLAS \citep{atlas}. Similar to UVOT, ZTF and ATLAS photometry were corrected for Galactic extinction.
The lightcurves are shown in Figure \ref{fig:lc}, where the top panel shows the UV/optical data and the bottom panel the X-ray observations.
The peak of the lightcurve is resolved in the ATLAS o-band lightcurve, and occurs $\approx$ 16 days after the first Gaia detection (MJD 58976). 

The Gaia lightcurve is retrieved from the Gaia Science alerts webpage\footnote{https://gsaweb.ast.cam.ac.uk/alerts} \citep{Hodgkin21}, illustrating a lack of optical variability in the host galaxy nucleus up to 2000 days before the discovery. 

\subsection{Optical spectroscopy of the host galaxy}
\begin{figure}
    \centering
    \includegraphics[width=\textwidth]{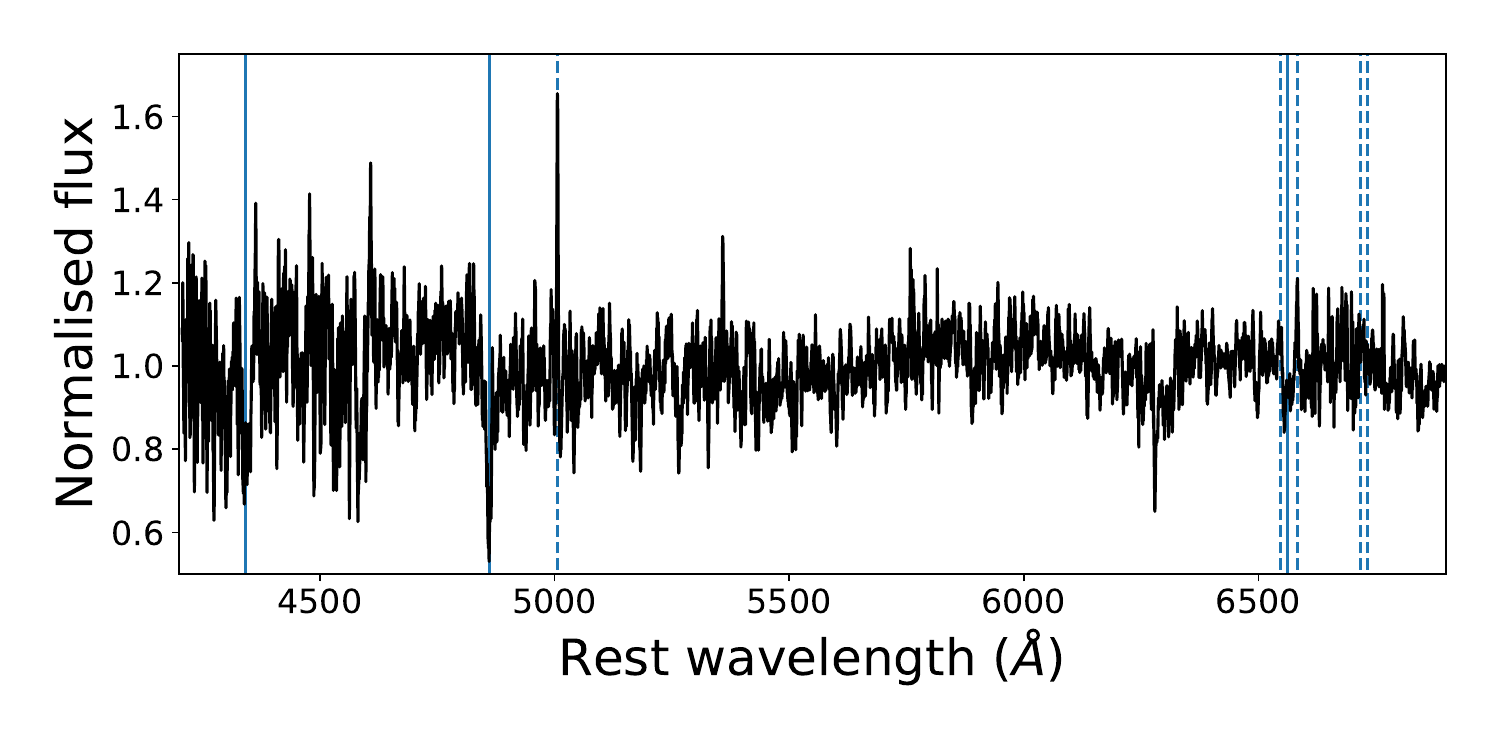}
    \caption{Continuum-normalized optical spectrum of the host galaxy nucleus of AT2020ksf, smoothed with a Gaussian kernel of 15 pixel width for presentation purposes. Solid lines indicate the Balmer transitions, while dashed lines indicate transitions typically seen in active and star forming galaxies.}
    \label{fig:hostspec}
\end{figure}

A spectrum of the host galaxy nucleus was taken with ESI on the Keck-II telescope in Mauna Kea, Hawaii on 2022 July 4. We observed the source using a 0.5 \arcsec slit and 600 seconds integration time. The resulting SNR is $\approx$6. Following the standard data reduction tasks, we normalize the spectrum to the continuum by dividing by a low order spline function fit to the continuum. A resampled, normalized version of the spectrum is shown in Figure \ref{fig:hostspec}. There are some emission lines typically observed in active and star forming galaxies, including O\,\textsc{iii} $\lambda 5007$ and the N\,\textsc{ii} doublet at $\lambda\lambda 6548, 6584$. The Balmer lines are seen in absorption, which could indicate the presence of a young stellar population. The presence of forbidden emission lines in combination with H Balmer lines in absorption is sometimes seen in post-starburst (E+A) galaxies, and is a trait seen in many TDE hosts galaxies \citep{French2016}. But to further quantify this would require a higher SNR spectrum that extends blueward to cover the H$\delta$ absorption line, and such a spectrum is not available.

Using the penalized pixel fitting routine \citep{Cappellari17} combined with the ELODIE stellar template library \citep{Prugniel2001, Prugniel2007}, we measure the velocity dispersion of the stellar absorption lines. We mask prominent emission lines (particularly, narrow host galaxy emission lines) during this process. Following \citet{Wevers17} we resample the spectrum within the errors and repeat the fitting procedure 1000 times, and take the mean and standard deviation as the velocity dispersion $\sigma$ and its uncertainty. We find $\sigma$ = 56$\pm$2 km s$^{-1}$, which translates into a black hole mass of log$_{10}$(M$_{\rm BH}$) = 5.2$\pm$0.46 M$_{\odot}$ using the M--$\sigma$ relation of \citet{Mcconnell13}, or alternatively log$_{10}$(M$_{\rm BH}$) = 6.1$\pm$0.35 M$_{\odot}$ using the \citet{Kormendy13} relation. Given the low SNR of the spectrum and the limited wavelength coverage, especially in the blue part of the spectrum, we interpret this result with the necessary caution. The resulting Eddington ratios that are derived from this measurement should be treated as rough estimations. 

\subsection{Limits on short term variability of the UFO}
\label{sec:intravar}
To further investigate potential variability on short ($\sim$ day) timescales, we further sub-divide the E34 data into two epochs, E3 and E4. The UFO is most significantly detected in epoch E3, and only marginally in epoch E4 ($\Delta \chi^2$ = 22 and 9 for E3 and E4, respectively). We note that the latter is significant only at the 2$\sigma$ level. We find values of the UFO parameters that are largely consistent within the (large) uncertainties (this is likely caused by the lower number of counts in the X-ray spectra, decreasing our sensitivity for constraining the UFO parameters). We hence do not find statistically significant evidence of further variability within these data. 
%If we assume UFO parameters as obtained from modeling spectrum E34, we can place a 1$\sigma$ upper limit on the column density in E4 of N$_{\rm H} < 8 \times 10^{22}$ cm$^{-2}$ (i.e. very similar to the value obtained from the UFO modeling of this epoch).

\bibliography{sample631}{}

\begin{thebibliography}{}
\expandafter\ifx\csname natexlab\endcsname\relax\def\natexlab#1{#1}\fi
\providecommand{\url}[1]{\href{#1}{#1}}
\providecommand{\dodoi}[1]{doi:~\href{http://doi.org/#1}{\nolinkurl{#1}}}
\providecommand{\doeprint}[1]{\href{http://ascl.net/#1}{\nolinkurl{http://ascl.net/#1}}}
\providecommand{\doarXiv}[1]{\href{https://arxiv.org/abs/#1}{\nolinkurl{https://arxiv.org/abs/#1}}}

\bibitem[{{Akaike}(1974)}]{Akaike1974}
{Akaike}, H. 1974, IEEE Transactions on Automatic Control, 19, 716

\bibitem[{{Alexander} {et~al.}(2017){Alexander}, {Wieringa}, {Berger},
  {Saxton}, \& {Komossa}}]{Alexander17}
{Alexander}, K.~D., {Wieringa}, M.~H., {Berger}, E., {Saxton}, R.~D., \&
  {Komossa}, S. 2017, \apj, 837, 153, \dodoi{10.3847/1538-4357/aa6192}

\bibitem[{{Alexander} {et~al.}(2021){Alexander}, {Velzen}, {Miller-Jones},
  {Anderson}, {Berger}, {Cendes}, {Chornock}, {Coppejans}, {Eftekhari},
  {Gezari}, {Goodwin}, {Hajela}, {Laskar}, {MacFadyen}, {Margutti}, {Pasham},
  {Ramirez-Ruiz}, \& {Saxton}}]{2021TNSAN..24....1A}
{Alexander}, K.~D., {Velzen}, S.~V., {Miller-Jones}, J., {et~al.} 2021,
  Transient Name Server AstroNote, 24, 1

\bibitem[{{Arnaud}(1996)}]{xspec}
{Arnaud}, K.~A. 1996, in Astronomical Society of the Pacific Conference Series,
  Vol. 101, Astronomical Data Analysis Software and Systems V, ed. G.~H.
  {Jacoby} \& J.~{Barnes}, 17

\bibitem[{{Bandopadhyay} {et~al.}(2023){Bandopadhyay}, {Fancher}, {Athian},
  {Indelicato}, {Kapalanga}, {Kumah}, {Paradiso}, {Todd}, {Coughlin}, \&
  {Nixon}}]{bandopadhyay23}
{Bandopadhyay}, A., {Fancher}, J., {Athian}, A., {et~al.} 2023, arXiv e-prints,
  arXiv:2310.11496, \dodoi{10.48550/arXiv.2310.11496}

\bibitem[{{Bellm} {et~al.}(2019){Bellm}, {Kulkarni}, {Graham}, {Dekany},
  {Smith}, {Riddle}, {Masci}, {Helou}, {Prince}, {Adams}, {Barbarino},
  {Barlow}, {Bauer}, {Beck}, {Belicki}, {Biswas}, {Blagorodnova}, {Bodewits},
  {Bolin}, {Brinnel}, {Brooke}, {Bue}, {Bulla}, {Burruss}, {Cenko}, {Chang},
  {Connolly}, {Coughlin}, {Cromer}, {Cunningham}, {De}, {Delacroix}, {Desai},
  {Duev}, {Eadie}, {Farnham}, {Feeney}, {Feindt}, {Flynn}, {Franckowiak},
  {Frederick}, {Fremling}, {Gal-Yam}, {Gezari}, {Giomi}, {Goldstein},
  {Golkhou}, {Goobar}, {Groom}, {Hacopians}, {Hale}, {Henning}, {Ho}, {Hover},
  {Howell}, {Hung}, {Huppenkothen}, {Imel}, {Ip}, {Ivezi{\'c}}, {Jackson},
  {Jones}, {Juric}, {Kasliwal}, {Kaspi}, {Kaye}, {Kelley}, {Kowalski},
  {Kramer}, {Kupfer}, {Landry}, {Laher}, {Lee}, {Lin}, {Lin}, {Lunnan},
  {Giomi}, {Mahabal}, {Mao}, {Miller}, {Monkewitz}, {Murphy}, {Ngeow},
  {Nordin}, {Nugent}, {Ofek}, {Patterson}, {Penprase}, {Porter}, {Rauch},
  {Rebbapragada}, {Reiley}, {Rigault}, {Rodriguez}, {van Roestel}, {Rusholme},
  {van Santen}, {Schulze}, {Shupe}, {Singer}, {Soumagnac}, {Stein}, {Surace},
  {Sollerman}, {Szkody}, {Taddia}, {Terek}, {Van Sistine}, {van Velzen},
  {Vestrand}, {Walters}, {Ward}, {Ye}, {Yu}, {Yan}, \& {Zolkower}}]{ztf}
{Bellm}, E.~C., {Kulkarni}, S.~R., {Graham}, M.~J., {et~al.} 2019, \pasp, 131,
  018002, \dodoi{10.1088/1538-3873/aaecbe}

\bibitem[{{Bright} {et~al.}(2018){Bright}, {Fender}, {Motta}, {Mooley},
  {Perrott}, {van Velzen}, {Carey}, {Hickish}, {Razavi-Ghods}, {Titterington},
  {Scott}, {Grainge}, {Scaife}, {Cantwell}, \& {Rumsey}}]{Bright18}
{Bright}, J.~S., {Fender}, R.~P., {Motta}, S.~E., {et~al.} 2018, \mnras, 475,
  4011, \dodoi{10.1093/mnras/sty077}

\bibitem[{{Bromberg} \& {Levinson}(2007)}]{bromberg07}
{Bromberg}, O., \& {Levinson}, A. 2007, \apj, 671, 678, \dodoi{10.1086/522668}

\bibitem[{Burnham \& Anderson(2002)}]{deltaaic}
Burnham, K., \& Anderson, D. 2002, Model selection and multimodel inference: a
  practical information-theoretic approach (Springer Verlag)

\bibitem[{{Burrows} {et~al.}(2005){Burrows}, {Hill}, {Nousek}, {Kennea},
  {Wells}, {Osborne}, {Abbey}, {Beardmore}, {Mukerjee}, {Short}, {Chincarini},
  {Campana}, {Citterio}, {Moretti}, {Pagani}, {Tagliaferri}, {Giommi},
  {Capalbi}, {Tamburelli}, {Angelini}, {Cusumano}, {Br{\"a}uninger}, {Burkert},
  \& {Hartner}}]{Burrows2005}
{Burrows}, D.~N., {Hill}, J.~E., {Nousek}, J.~A., {et~al.} 2005, \ssr, 120,
  165, \dodoi{10.1007/s11214-005-5097-2}

\bibitem[{{Cannizzo} {et~al.}(1990){Cannizzo}, {Lee}, \&
  {Goodman}}]{cannizzo90}
{Cannizzo}, J.~K., {Lee}, H.~M., \& {Goodman}, J. 1990, \apj, 351, 38,
  \dodoi{10.1086/168442}

\bibitem[{{Cappellari}(2017)}]{Cappellari17}
{Cappellari}, M. 2017, \mnras, 466, 798, \dodoi{10.1093/mnras/stw3020}

\bibitem[{{Cappi}(2006)}]{Cappi06}
{Cappi}, M. 2006, Astronomische Nachrichten, 327, 1012,
  \dodoi{10.1002/asna.200610639}

\bibitem[{{Cash}(1979)}]{cstat}
{Cash}, W. 1979, \apj, 228, 939, \dodoi{10.1086/156922}

\bibitem[{{Cendes} {et~al.}(2023){Cendes}, {Berger}, {Alexander}, {Chornock},
  {Margutti}, {Metzger}, {Wieringa}, {Bietenholz}, {Hajela}, {Laskar}, {Stroh},
  \& {Terreran}}]{Cendes23}
{Cendes}, Y., {Berger}, E., {Alexander}, K.~D., {et~al.} 2023, arXiv e-prints,
  arXiv:2308.13595, \dodoi{10.48550/arXiv.2308.13595}

\bibitem[{{Chan} {et~al.}(2022){Chan}, {Piran}, \& {Krolik}}]{Chan2022}
{Chan}, C.-H., {Piran}, T., \& {Krolik}, J.~H. 2022, \apj, 933, 81,
  \dodoi{10.3847/1538-4357/ac68f3}

\bibitem[{{Coughlin} \& {Begelman}(2014)}]{coughlin14}
{Coughlin}, E.~R., \& {Begelman}, M.~C. 2014, \apj, 781, 82,
  \dodoi{10.1088/0004-637X/781/2/82}

\bibitem[{{Coughlin} \& {Begelman}(2020)}]{coughlin20}
---. 2020, \mnras, 499, 3158, \dodoi{10.1093/mnras/staa3026}

\bibitem[{{Coughlin} \& {Nixon}(2019)}]{coughlin19}
{Coughlin}, E.~R., \& {Nixon}, C.~J. 2019, \apjl, 883, L17,
  \dodoi{10.3847/2041-8213/ab412d}

\bibitem[{{Coughlin} \& {Nixon}(2022)}]{coughlin22}
---. 2022, \mnras, 517, L26, \dodoi{10.1093/mnrasl/slac106}

\bibitem[{{Dai} {et~al.}(2018){Dai}, {McKinney}, {Roth}, {Ramirez-Ruiz}, \&
  {Miller}}]{Dai2018}
{Dai}, L., {McKinney}, J.~C., {Roth}, N., {Ramirez-Ruiz}, E., \& {Miller},
  M.~C. 2018, \apjl, 859, L20, \dodoi{10.3847/2041-8213/aab429}

\bibitem[{{Dexter} \& {Begelman}(2019)}]{dexter19}
{Dexter}, J., \& {Begelman}, M.~C. 2019, \mnras, 483, L17,
  \dodoi{10.1093/mnrasl/sly213}

\bibitem[{{Ellison} {et~al.}(2018){Ellison}, {Catinella}, \&
  {Cortese}}]{Ellison2018}
{Ellison}, S.~L., {Catinella}, B., \& {Cortese}, L. 2018, \mnras, 478, 3447,
  \dodoi{10.1093/mnras/sty1247}

\bibitem[{{Evans} {et~al.}(2009){Evans}, {Beardmore}, {Page}, {Osborne},
  {O'Brien}, {Willingale}, {Starling}, {Burrows}, {Godet}, {Vetere}, {Racusin},
  {Goad}, {Wiersema}, {Angelini}, {Capalbi}, {Chincarini}, {Gehrels}, {Kennea},
  {Margutti}, {Morris}, {Mountford}, {Pagani}, {Perri}, {Romano}, \&
  {Tanvir}}]{Evans2009}
{Evans}, P.~A., {Beardmore}, A.~P., {Page}, K.~L., {et~al.} 2009, \mnras, 397,
  1177, \dodoi{10.1111/j.1365-2966.2009.14913.x}

\bibitem[{{Eyles-Ferris} {et~al.}(2022){Eyles-Ferris}, {Starling}, {O'Brien},
  {Nixon}, \& {Coughlin}}]{eyles22}
{Eyles-Ferris}, R.~A.~J., {Starling}, R.~L.~C., {O'Brien}, P.~T., {Nixon},
  C.~J., \& {Coughlin}, E.~R. 2022, \mnras, 517, 6013,
  \dodoi{10.1093/mnras/stac3073}

\bibitem[{{Fangyi} {et~al.}(2023){Fangyi}, {Hu}, {Price}, \& {Mandel}}]{fangyi}
{Fangyi}, {Hu}, {Price}, D.~J., \& {Mandel}, I. 2023, arXiv e-prints,
  arXiv:2312.03210, \dodoi{10.48550/arXiv.2312.03210}

\bibitem[{{French} {et~al.}(2016){French}, {Arcavi}, \&
  {Zabludoff}}]{French2016}
{French}, K.~D., {Arcavi}, I., \& {Zabludoff}, A. 2016, \apjl, 818, L21,
  \dodoi{10.3847/2041-8205/818/1/L21}

\bibitem[{{French} {et~al.}(2020){French}, {Wevers}, {Law-Smith}, {Graur}, \&
  {Zabludoff}}]{French2020}
{French}, K.~D., {Wevers}, T., {Law-Smith}, J., {Graur}, O., \& {Zabludoff},
  A.~I. 2020, \ssr, 216, 32, \dodoi{10.1007/s11214-020-00657-y}

\bibitem[{{French} {et~al.}(2018){French}, {Yang}, {Zabludoff}, \&
  {Tremonti}}]{French18}
{French}, K.~D., {Yang}, Y., {Zabludoff}, A.~I., \& {Tremonti}, C.~A. 2018,
  \apj, 862, 2, \dodoi{10.3847/1538-4357/aacb2d}

\bibitem[{{Gabriel} {et~al.}(2004){Gabriel}, {Denby}, {Fyfe}, {Hoar}, {Ibarra},
  {Ojero}, {Osborne}, {Saxton}, {Lammers}, \& {Vacanti}}]{Gabriel_04}
{Gabriel}, C., {Denby}, M., {Fyfe}, D.~J., {et~al.} 2004, in Astronomical
  Society of the Pacific Conference Series, Vol. 314, Astronomical Data
  Analysis Software and Systems (ADASS) XIII, ed. F.~{Ochsenbein}, M.~G.
  {Allen}, \& D.~{Egret}, 759

\bibitem[{{Gendreau} {et~al.}(2016){Gendreau}, {Arzoumanian}, {Adkins},
  {Albert}, {Anders}, {Aylward}, {Baker}, {Balsamo}, {Bamford}, {Benegalrao},
  {Berry}, {Bhalwani}, {Black}, {Blaurock}, {Bronke}, {Brown}, {Budinoff},
  {Cantwell}, {Cazeau}, {Chen}, {Clement}, {Colangelo}, {Coleman},
  {Coopersmith}, {Dehaven}, {Doty}, {Egan}, {Enoto}, {Fan}, {Ferro}, {Foster},
  {Galassi}, {Gallo}, {Green}, {Grosh}, {Ha}, {Hasouneh}, {Heefner}, {Hestnes},
  {Hoge}, {Jacobs}, {J{\o}rgensen}, {Kaiser}, {Kellogg}, {Kenyon}, {Koenecke},
  {Kozon}, {LaMarr}, {Lambertson}, {Larson}, {Lentine}, {Lewis}, {Lilly},
  {Liu}, {Malonis}, {Manthripragada}, {Markwardt}, {Matonak}, {Mcginnis},
  {Miller}, {Mitchell}, {Mitchell}, {Mohammed}, {Monroe}, {Montt de Garcia},
  {Mul{\'e}}, {Nagao}, {Ngo}, {Norris}, {Norwood}, {Novotka}, {Okajima},
  {Olsen}, {Onyeachu}, {Orosco}, {Peterson}, {Pevear}, {Pham}, {Pollard},
  {Pope}, {Powers}, {Powers}, {Price}, {Prigozhin}, {Ramirez}, {Reid},
  {Remillard}, {Rogstad}, {Rosecrans}, {Rowe}, {Sager}, {Sanders}, {Savadkin},
  {Saylor}, {Schaeffer}, {Schweiss}, {Semper}, {Serlemitsos}, {Shackelford},
  {Soong}, {Struebel}, {Vezie}, {Villasenor}, {Winternitz}, {Wofford},
  {Wright}, {Yang}, \& {Yu}}]{keith}
{Gendreau}, K.~C., {Arzoumanian}, Z., {Adkins}, P.~W., {et~al.} 2016, in
  Society of Photo-Optical Instrumentation Engineers (SPIE) Conference Series,
  Vol. 9905, Space Telescopes and Instrumentation 2016: Ultraviolet to Gamma
  Ray, ed. J.-W.~A. {den Herder}, T.~{Takahashi}, \& M.~{Bautz}, 99051H,
  \dodoi{10.1117/12.2231304}

\bibitem[{{Gezari} {et~al.}(2017){Gezari}, {Cenko}, \& {Arcavi}}]{Gezari2017}
{Gezari}, S., {Cenko}, S.~B., \& {Arcavi}, I. 2017, \apjl, 851, L47,
  \dodoi{10.3847/2041-8213/aaa0c2}

\bibitem[{{Gilfanov} {et~al.}(2020){Gilfanov}, {Sazonov}, {Sunyaev},
  {Medvedev}, {Khorunzhev}, {Semena}, {Yao}, {Kulkarni}, {Gezari}, \& {van
  Velzen}}]{2020ATel14246....1G}
{Gilfanov}, M., {Sazonov}, S., {Sunyaev}, R., {et~al.} 2020, The Astronomer's
  Telegram, 14246, 1

\bibitem[{{Gu{\'e}pin} {et~al.}(2022){Gu{\'e}pin}, {Kotera}, \&
  {Oikonomou}}]{guepin}
{Gu{\'e}pin}, C., {Kotera}, K., \& {Oikonomou}, F. 2022, Nature Reviews
  Physics, 4, 697, \dodoi{10.1038/s42254-022-00504-9}

\bibitem[{{Guillochon} \& {Ramirez-Ruiz}(2013)}]{guillochon13}
{Guillochon}, J., \& {Ramirez-Ruiz}, E. 2013, \apj, 767, 25,
  \dodoi{10.1088/0004-637X/767/1/25}

\bibitem[{{Guillochon} \& {Ramirez-Ruiz}(2015)}]{guillochon15}
---. 2015, \apj, 809, 166, \dodoi{10.1088/0004-637X/809/2/166}

\bibitem[{{Guolo} {et~al.}(2023){Guolo}, {Gezari}, {Yao}, {van Velzen},
  {Hammerstein}, {Cenko}, \& {Tokayer}}]{Guolo23}
{Guolo}, M., {Gezari}, S., {Yao}, Y., {et~al.} 2023, arXiv e-prints,
  arXiv:2308.13019, \dodoi{10.48550/arXiv.2308.13019}

\bibitem[{{Hammerstein} {et~al.}(2023){Hammerstein}, {van Velzen}, {Gezari},
  {Cenko}, {Yao}, {Ward}, {Frederick}, {Villanueva}, {Somalwar}, {Graham},
  {Kulkarni}, {Stern}, {Andreoni}, {Bellm}, {Dekany}, {Dhawan}, {Drake},
  {Fremling}, {Gatkine}, {Groom}, {Ho}, {Kasliwal}, {Karambelkar}, {Kool},
  {Masci}, {Medford}, {Perley}, {Purdum}, {van Roestel}, {Sharma}, {Sollerman},
  {Taggart}, \& {Yan}}]{Hammerstein2023}
{Hammerstein}, E., {van Velzen}, S., {Gezari}, S., {et~al.} 2023, \apj, 942, 9,
  \dodoi{10.3847/1538-4357/aca283}

\bibitem[{{Harrison} {et~al.}(2018){Harrison}, {Costa}, {Tadhunter},
  {Fl{\"u}tsch}, {Kakkad}, {Perna}, \& {Vietri}}]{Harrison2018}
{Harrison}, C.~M., {Costa}, T., {Tadhunter}, C.~N., {et~al.} 2018, Nature
  Astronomy, 2, 198, \dodoi{10.1038/s41550-018-0403-6}

\bibitem[{{Hawley} {et~al.}(2011){Hawley}, {Guan}, \& {Krolik}}]{hawley11}
{Hawley}, J.~F., {Guan}, X., \& {Krolik}, J.~H. 2011, \apj, 738, 84,
  \dodoi{10.1088/0004-637X/738/1/84}

\bibitem[{{HI4PI Collaboration} {et~al.}(2016){HI4PI Collaboration}, {Ben
  Bekhti}, {Fl{\"o}er}, {Keller}, {Kerp}, {Lenz}, {Winkel}, {Bailin},
  {Calabretta}, {Dedes}, {Ford}, {Gibson}, {Haud}, {Janowiecki}, {Kalberla},
  {Lockman}, {McClure-Griffiths}, {Murphy}, {Nakanishi}, {Pisano}, \&
  {Staveley-Smith}}]{hi4pi}
{HI4PI Collaboration}, {Ben Bekhti}, N., {Fl{\"o}er}, L., {et~al.} 2016, \aap,
  594, A116, \dodoi{10.1051/0004-6361/201629178}

\bibitem[{{Hinkle} {et~al.}(2021){Hinkle}, {Holoien}, {Auchettl}, {Shappee},
  {Neustadt}, {Payne}, {Brown}, {Kochanek}, {Stanek}, {Graham}, {Tucker}, {Do},
  {Anderson}, {Bose}, {Chen}, {Coulter}, {Dimitriadis}, {Dong}, {Foley},
  {Huber}, {Hung}, {Kilpatrick}, {Pignata}, {Piro}, {Rojas-Bravo}, {Siebert},
  {Stalder}, {Thompson}, {Tonry}, {Vallely}, \& {Wisniewski}}]{Hinkle2021}
{Hinkle}, J.~T., {Holoien}, T.~W.~S., {Auchettl}, K., {et~al.} 2021, \mnras,
  500, 1673, \dodoi{10.1093/mnras/staa3170}

\bibitem[{{Hodgkin} {et~al.}(2021){Hodgkin}, {Harrison}, {Breedt}, {Wevers},
  {Rixon}, {Delgado}, {Yoldas}, {Kostrzewa-Rutkowska}, {Wyrzykowski}, {van
  Leeuwen}, {Blagorodnova}, {Campbell}, {Eappachen}, {Fraser}, {Ihanec},
  {Koposov}, {Kruszy{\'n}ska}, {Marton}, {Rybicki}, {Brown}, {Burgess},
  {Busso}, {Cowell}, {De Angeli}, {Diener}, {Evans}, {Gilmore}, {Holland},
  {Jonker}, {van Leeuwen}, {Mignard}, {Osborne}, {Portell}, {Prusti},
  {Richards}, {Riello}, {Seabroke}, {Walton}, {{\'A}brah{\'a}m}, {Altavilla},
  {Baker}, {Bastian}, {O'Brien}, {de Bruijne}, {Butterley}, {Carrasco},
  {Casta{\~n}eda}, {Clark}, {Clementini}, {Copperwheat}, {Cropper},
  {Damljanovic}, {Davidson}, {Davis}, {Dennefeld}, {Dhillon}, {Dolding},
  {Dominik}, {Esquej}, {Eyer}, {Fabricius}, {Fridman}, {Froebrich}, {Garralda},
  {Gomboc}, {Gonz{\'a}lez-Vidal}, {Guerra}, {Hambly}, {Hardy}, {Holl},
  {Hourihane}, {Japelj}, {Kann}, {Kiss}, {Knigge}, {Kolb}, {Komossa},
  {K{\'o}sp{\'a}l}, {Kov{\'a}cs}, {Kun}, {Leto}, {Lewis}, {Littlefair},
  {Mahabal}, {Mundell}, {Nagy}, {Padeletti}, {Palaversa}, {Pigulski},
  {Pretorius}, {van Reeven}, {Ribeiro}, {Roelens}, {Rowell}, {Schartel},
  {Scholz}, {Schwope}, {Sip{\H{o}}cz}, {Smartt}, {Smith}, {Serraller},
  {Steeghs}, {Sullivan}, {Szabados}, {Szegedi-Elek}, {Tisserand}, {Tomasella},
  {van Velzen}, {Whitelock}, {Wilson}, \& {Young}}]{Hodgkin21}
{Hodgkin}, S.~T., {Harrison}, D.~L., {Breedt}, E., {et~al.} 2021, \aap, 652,
  A76, \dodoi{10.1051/0004-6361/202140735}

\bibitem[{{Jansen} {et~al.}(2001){Jansen}, {Lumb}, {Altieri}, {Clavel}, {Ehle},
  {Erd}, {Gabriel}, {Guainazzi}, {Gondoin}, {Much}, {Munoz}, {Santos},
  {Schartel}, {Texier}, \& {Vacanti}}]{xmm}
{Jansen}, F., {Lumb}, D., {Altieri}, B., {et~al.} 2001, \aap, 365, L1,
  \dodoi{10.1051/0004-6361:20000036}

\bibitem[{{Jiang} {et~al.}(2021){Jiang}, {Wang}, {Hu}, {Sun}, {Dou}, \&
  {Xiao}}]{Jiang2021}
{Jiang}, N., {Wang}, T., {Hu}, X., {et~al.} 2021, \apj, 911, 31,
  \dodoi{10.3847/1538-4357/abe772}

\bibitem[{{Johnson} {et~al.}(2021){Johnson}, {Leja}, {Conroy}, \&
  {Speagle}}]{Johnson_21}
{Johnson}, B.~D., {Leja}, J., {Conroy}, C., \& {Speagle}, J.~S. 2021, \apjs,
  254, 22, \dodoi{10.3847/1538-4365/abef67}

\bibitem[{{Kaastra} \& {Bleeker}(2016)}]{kaastra}
{Kaastra}, J.~S., \& {Bleeker}, J.~A.~M. 2016, \aap, 587, A151,
  \dodoi{10.1051/0004-6361/201527395}

\bibitem[{{Kajava} {et~al.}(2020){Kajava}, {Giustini}, {Saxton}, \&
  {Miniutti}}]{Kajava2020}
{Kajava}, J. J.~E., {Giustini}, M., {Saxton}, R.~D., \& {Miniutti}, G. 2020,
  \aap, 639, A100, \dodoi{10.1051/0004-6361/202038165}

\bibitem[{{Kara} {et~al.}(2018){Kara}, {Dai}, {Reynolds}, \&
  {Kallman}}]{Kara18}
{Kara}, E., {Dai}, L., {Reynolds}, C.~S., \& {Kallman}, T. 2018, \mnras, 474,
  3593, \dodoi{10.1093/mnras/stx3004}

\bibitem[{{King} \& {Pounds}(2015)}]{King2015}
{King}, A., \& {Pounds}, K. 2015, \araa, 53, 115,
  \dodoi{10.1146/annurev-astro-082214-122316}

\bibitem[{{King} {et~al.}(2007){King}, {Pringle}, \& {Livio}}]{king07}
{King}, A.~R., {Pringle}, J.~E., \& {Livio}, M. 2007, \mnras, 376, 1740,
  \dodoi{10.1111/j.1365-2966.2007.11556.x}

\bibitem[{{Kobayashi} {et~al.}(2018){Kobayashi}, {Ohsuga}, {Takahashi},
  {Kawashima}, {Asahina}, {Takeuchi}, \& {Mineshige}}]{Kobayashi2018}
{Kobayashi}, H., {Ohsuga}, K., {Takahashi}, H.~R., {et~al.} 2018, \pasj, 70,
  22, \dodoi{10.1093/pasj/psx157}

\bibitem[{{Kochanek}(1994)}]{Kochanek94}
{Kochanek}, C.~S. 1994, \apj, 422, 508, \dodoi{10.1086/173745}

\bibitem[{{Kohler} {et~al.}(2012){Kohler}, {Begelman}, \&
  {Beckwith}}]{kohler12}
{Kohler}, S., {Begelman}, M.~C., \& {Beckwith}, K. 2012, \mnras, 422, 2282,
  \dodoi{10.1111/j.1365-2966.2012.20776.x}

\bibitem[{{Kormendy} \& {Ho}(2013)}]{Kormendy13}
{Kormendy}, J., \& {Ho}, L.~C. 2013, \araa, 51, 511,
  \dodoi{10.1146/annurev-astro-082708-101811}

\bibitem[{{Kosec} {et~al.}(2023){Kosec}, {Pasham}, {Kara}, \&
  {Tombesi}}]{Kosec23}
{Kosec}, P., {Pasham}, D., {Kara}, E., \& {Tombesi}, F. 2023, arXiv e-prints,
  arXiv:2308.05250, \dodoi{10.48550/arXiv.2308.05250}

\bibitem[{{Kosec} {et~al.}(2018){Kosec}, {Pinto}, {Walton}, {Fabian},
  {Bachetti}, {Brightman}, {F{\"u}rst}, \& {Grefenstette}}]{Kosec2018}
{Kosec}, P., {Pinto}, C., {Walton}, D.~J., {et~al.} 2018, \mnras, 479, 3978,
  \dodoi{10.1093/mnras/sty1626}

\bibitem[{{Law-Smith} {et~al.}(2020){Law-Smith}, {Coulter}, {Guillochon},
  {Mockler}, \& {Ramirez-Ruiz}}]{lawsmith20}
{Law-Smith}, J. A.~P., {Coulter}, D.~A., {Guillochon}, J., {Mockler}, B., \&
  {Ramirez-Ruiz}, E. 2020, \apj, 905, 141, \dodoi{10.3847/1538-4357/abc489}

\bibitem[{{Lin} {et~al.}(2015){Lin}, {Maksym}, {Irwin}, {Komossa}, {Webb},
  {Godet}, {Barret}, {Grupe}, \& {Gwyn}}]{Lin2015}
{Lin}, D., {Maksym}, P.~W., {Irwin}, J.~A., {et~al.} 2015, \apj, 811, 43,
  \dodoi{10.1088/0004-637X/811/1/43}

\bibitem[{{Lixin Dai} {et~al.}(2021){Lixin Dai}, {Lodato}, \&
  {Cheng}}]{Dai2021}
{Lixin Dai}, J., {Lodato}, G., \& {Cheng}, R.~M. 2021, arXiv e-prints,
  arXiv:2101.05195, \dodoi{10.48550/arXiv.2101.05195}

\bibitem[{{Lu} \& {Bonnerot}(2020)}]{Lu2020}
{Lu}, W., \& {Bonnerot}, C. 2020, \mnras, 492, 686,
  \dodoi{10.1093/mnras/stz3405}

\bibitem[{{Mainetti} {et~al.}(2017){Mainetti}, {Lupi}, {Campana}, {Colpi},
  {Coughlin}, {Guillochon}, \& {Ramirez-Ruiz}}]{mainetti17}
{Mainetti}, D., {Lupi}, A., {Campana}, S., {et~al.} 2017, \aap, 600, A124,
  \dodoi{10.1051/0004-6361/201630092}

\bibitem[{{Masci} {et~al.}(2019){Masci}, {Laher}, {Rusholme}, {Shupe}, {Groom},
  {Surace}, {Jackson}, {Monkewitz}, {Beck}, {Flynn}, {Terek}, {Landry},
  {Hacopians}, {Desai}, {Howell}, {Brooke}, {Imel}, {Wachter}, {Ye}, {Lin},
  {Cenko}, {Cunningham}, {Rebbapragada}, {Bue}, {Miller}, {Mahabal}, {Bellm},
  {Patterson}, {Juri{\'c}}, {Golkhou}, {Ofek}, {Walters}, {Graham}, {Kasliwal},
  {Dekany}, {Kupfer}, {Burdge}, {Cannella}, {Barlow}, {Van Sistine}, {Giomi},
  {Fremling}, {Blagorodnova}, {Levitan}, {Riddle}, {Smith}, {Helou}, {Prince},
  \& {Kulkarni}}]{Masci2019}
{Masci}, F.~J., {Laher}, R.~R., {Rusholme}, B., {et~al.} 2019, \pasp, 131,
  018003, \dodoi{10.1088/1538-3873/aae8ac}

\bibitem[{{McConnell} \& {Ma}(2013)}]{Mcconnell13}
{McConnell}, N.~J., \& {Ma}, C.-P. 2013, \apj, 764, 184,
  \dodoi{10.1088/0004-637X/764/2/184}

\bibitem[{{Metzger}(2022)}]{Metzger22}
{Metzger}, B.~D. 2022, \apjl, 937, L12, \dodoi{10.3847/2041-8213/ac90ba}

\bibitem[{{Miles} {et~al.}(2020){Miles}, {Coughlin}, \& {Nixon}}]{miles2020}
{Miles}, P.~R., {Coughlin}, E.~R., \& {Nixon}, C.~J. 2020, \apj, 899, 36,
  \dodoi{10.3847/1538-4357/ab9c9f}

\bibitem[{{Miller} {et~al.}(2015){Miller}, {Kaastra}, {Miller}, {Reynolds},
  {Brown}, {Cenko}, {Drake}, {Gezari}, {Guillochon}, {Gultekin}, {Irwin},
  {Levan}, {Maitra}, {Maksym}, {Mushotzky}, {O'Brien}, {Paerels}, {de Plaa},
  {Ramirez-Ruiz}, {Strohmayer}, \& {Tanvir}}]{Miller2015}
{Miller}, J.~M., {Kaastra}, J.~S., {Miller}, M.~C., {et~al.} 2015, \nat, 526,
  542, \dodoi{10.1038/nature15708}

\bibitem[{{Mummery} {et~al.}(2023){Mummery}, {Wevers}, {Saxton}, \&
  {Pasham}}]{Mummery23}
{Mummery}, A., {Wevers}, T., {Saxton}, R., \& {Pasham}, D. 2023, \mnras, 519,
  5828, \dodoi{10.1093/mnras/stac3798}

\bibitem[{{Murase} {et~al.}(2020){Murase}, {Kimura}, {Zhang}, {Oikonomou}, \&
  {Petropoulou}}]{murase}
{Murase}, K., {Kimura}, S.~S., {Zhang}, B.~T., {Oikonomou}, F., \&
  {Petropoulou}, M. 2020, \apj, 902, 108, \dodoi{10.3847/1538-4357/abb3c0}

\bibitem[{{Nardini} {et~al.}(2015){Nardini}, {Reeves}, {Gofford}, {Harrison},
  {Risaliti}, {Braito}, {Costa}, {Matzeu}, {Walton}, {Behar}, {Boggs},
  {Christensen}, {Craig}, {Hailey}, {Matt}, {Miller}, {O'Brien}, {Stern},
  {Turner}, \& {Ward}}]{Nardini2015}
{Nardini}, E., {Reeves}, J.~N., {Gofford}, J., {et~al.} 2015, Science, 347,
  860, \dodoi{10.1126/science.1259202}

\bibitem[{{Nixon} {et~al.}(2021){Nixon}, {Coughlin}, \& {Miles}}]{nixon21}
{Nixon}, C.~J., {Coughlin}, E.~R., \& {Miles}, P.~R. 2021, \apj, 922, 168,
  \dodoi{10.3847/1538-4357/ac1bb8}

\bibitem[{{Parkinson} {et~al.}(2022){Parkinson}, {Knigge}, {Matthews}, {Long},
  {Higginbottom}, {Sim}, \& {Mangham}}]{Parkinson22}
{Parkinson}, E.~J., {Knigge}, C., {Matthews}, J.~H., {et~al.} 2022, \mnras,
  510, 5426, \dodoi{10.1093/mnras/stac027}

\bibitem[{{Pasham} {et~al.}(2023){Pasham}, {Lucchini}, {Laskar}, {Gompertz},
  {Srivastav}, {Nicholl}, {Smartt}, {Miller-Jones}, {Alexander}, {Fender},
  {Smith}, {Fulton}, {Dewangan}, {Gendreau}, {Coughlin}, {Rhodes}, {Horesh},
  {van Velzen}, {Sfaradi}, {Guolo}, {Castro Segura}, {Aamer}, {Anderson},
  {Arcavi}, {Brennan}, {Chambers}, {Charalampopoulos}, {Chen}, {Clocchiatti},
  {de Boer}, {Dennefeld}, {Ferrara}, {Galbany}, {Gao}, {Gillanders}, {Goodwin},
  {Gromadzki}, {Huber}, {Jonker}, {Joshi}, {Kara}, {Killestein}, {Kosec},
  {Kocevski}, {Leloudas}, {Lin}, {Margutti}, {Mattila}, {Moore},
  {M{\"u}ller-Bravo}, {Ngeow}, {Oates}, {Onori}, {Pan}, {Perez-Torres}, {Rani},
  {Remillard}, {Ridley}, {Schulze}, {Sheng}, {Shingles}, {Smith}, {Steiner},
  {Wainscoat}, {Wevers}, \& {Yang}}]{2022cmc}
{Pasham}, D.~R., {Lucchini}, M., {Laskar}, T., {et~al.} 2023, Nature Astronomy,
  7, 88, \dodoi{10.1038/s41550-022-01820-x}

\bibitem[{{Pinto} {et~al.}(2018){Pinto}, {Alston}, {Parker}, {Fabian}, {Gallo},
  {Buisson}, {Walton}, {Kara}, {Jiang}, {Lohfink}, \& {Reynolds}}]{Pinto2018}
{Pinto}, C., {Alston}, W., {Parker}, M.~L., {et~al.} 2018, \mnras, 476, 1021,
  \dodoi{10.1093/mnras/sty231}

\bibitem[{{Pinto} {et~al.}(2020){Pinto}, {Mehdipour}, {Walton}, {Middleton},
  {Roberts}, {Fabian}, {Guainazzi}, {Soria}, {Kosec}, \& {Ness}}]{Pinto2020}
{Pinto}, C., {Mehdipour}, M., {Walton}, D.~J., {et~al.} 2020, \mnras, 491,
  5702, \dodoi{10.1093/mnras/stz3392}

\bibitem[{{Pinto} {et~al.}(2021){Pinto}, {Soria}, {Walton}, {D'A{\`\i}},
  {Pintore}, {Kosec}, {Alston}, {Fuerst}, {Middleton}, {Roberts}, {Del Santo},
  {Barret}, {Ambrosi}, {Robba}, {Earnshaw}, \& {Fabian}}]{Pinto2021}
{Pinto}, C., {Soria}, R., {Walton}, D.~J., {et~al.} 2021, \mnras, 505, 5058,
  \dodoi{10.1093/mnras/stab1648}

\bibitem[{{Piran} {et~al.}(2015){Piran}, {Svirski}, {Krolik}, {Cheng}, \&
  {Shiokawa}}]{Piran15}
{Piran}, T., {Svirski}, G., {Krolik}, J., {Cheng}, R.~M., \& {Shiokawa}, H.
  2015, \apj, 806, 164, \dodoi{10.1088/0004-637X/806/2/164}

\bibitem[{{Pounds} {et~al.}(2003){Pounds}, {Reeves}, {King}, {Page}, {O'Brien},
  \& {Turner}}]{Pounds03}
{Pounds}, K.~A., {Reeves}, J.~N., {King}, A.~R., {et~al.} 2003, \mnras, 345,
  705, \dodoi{10.1046/j.1365-8711.2003.07006.x}

\bibitem[{{Predehl} {et~al.}(2021){Predehl}, {Andritschke}, {Arefiev},
  {Babyshkin}, {Batanov}, {Becker}, {B{\"o}hringer}, {Bogomolov}, {Boller},
  {Borm}, {Bornemann}, {Br{\"a}uninger}, {Br{\"u}ggen}, {Brunner}, {Brusa},
  {Bulbul}, {Buntov}, {Burwitz}, {Burkert}, {Clerc}, {Churazov}, {Coutinho},
  {Dauser}, {Dennerl}, {Doroshenko}, {Eder}, {Emberger}, {Eraerds},
  {Finoguenov}, {Freyberg}, {Friedrich}, {Friedrich}, {F{\"u}rmetz},
  {Georgakakis}, {Gilfanov}, {Granato}, {Grossberger}, {Gueguen}, {Gureev},
  {Haberl}, {H{\"a}lker}, {Hartner}, {Hasinger}, {Huber}, {Ji}, {Kienlin},
  {Kink}, {Korotkov}, {Kreykenbohm}, {Lamer}, {Lomakin}, {Lapshov}, {Liu},
  {Maitra}, {Meidinger}, {Menz}, {Merloni}, {Mernik}, {Mican}, {Mohr},
  {M{\"u}ller}, {Nandra}, {Nazarov}, {Pacaud}, {Pavlinsky}, {Perinati},
  {Pfeffermann}, {Pietschner}, {Ramos-Ceja}, {Rau}, {Reiffers}, {Reiprich},
  {Robrade}, {Salvato}, {Sanders}, {Santangelo}, {Sasaki}, {Scheuerle},
  {Schmid}, {Schmitt}, {Schwope}, {Shirshakov}, {Steinmetz}, {Stewart},
  {Str{\"u}der}, {Sunyaev}, {Tenzer}, {Tiedemann}, {Tr{\"u}mper}, {Voron},
  {Weber}, {Wilms}, \& {Yaroshenko}}]{Predehl21}
{Predehl}, P., {Andritschke}, R., {Arefiev}, V., {et~al.} 2021, \aap, 647, A1,
  \dodoi{10.1051/0004-6361/202039313}

\bibitem[{{Prugniel} \& {Soubiran}(2001)}]{Prugniel2001}
{Prugniel}, P., \& {Soubiran}, C. 2001, \aap, 369, 1048,
  \dodoi{10.1051/0004-6361:20010163}

\bibitem[{{Prugniel} {et~al.}(2007){Prugniel}, {Soubiran}, {Koleva}, \& {Le
  Borgne}}]{Prugniel2007}
{Prugniel}, P., {Soubiran}, C., {Koleva}, M., \& {Le Borgne}, D. 2007, VizieR
  Online Data Catalog, III/251

\bibitem[{{Rees}(1988)}]{Rees1988}
{Rees}, M.~J. 1988, \nat, 333, 523, \dodoi{10.1038/333523a0}

\bibitem[{{Remillard} {et~al.}(2022){Remillard}, {Loewenstein}, {Steiner},
  {Prigozhin}, {LaMarr}, {Enoto}, {Gendreau}, {Arzoumanian}, {Markwardt},
  {Basak}, {Stevens}, {Ray}, {Altamirano}, \& {Buisson}}]{3c50}
{Remillard}, R.~A., {Loewenstein}, M., {Steiner}, J.~F., {et~al.} 2022, \aj,
  163, 130, \dodoi{10.3847/1538-3881/ac4ae6}

\bibitem[{{Roming} {et~al.}(2005){Roming}, {Kennedy}, {Mason}, {Nousek}, {Ahr},
  {Bingham}, {Broos}, {Carter}, {Hancock}, \& {Huckle}}]{Roming2005}
{Roming}, P. W.~A., {Kennedy}, T.~E., {Mason}, K.~O., {et~al.} 2005, \ssr, 120,
  95, \dodoi{10.1007/s11214-005-5095-4}

\bibitem[{{Roth} {et~al.}(2016){Roth}, {Kasen}, {Guillochon}, \&
  {Ramirez-Ruiz}}]{roth16}
{Roth}, N., {Kasen}, D., {Guillochon}, J., \& {Ramirez-Ruiz}, E. 2016, \apj,
  827, 3, \dodoi{10.3847/0004-637X/827/1/3}

\bibitem[{{Ryu} {et~al.}(2020){Ryu}, {Krolik}, {Piran}, \& {Noble}}]{ryu2020}
{Ryu}, T., {Krolik}, J., {Piran}, T., \& {Noble}, S.~C. 2020, \apj, 904, 98,
  \dodoi{10.3847/1538-4357/abb3cf}

\bibitem[{{Ryu} {et~al.}(2023){Ryu}, {Krolik}, {Piran}, {Noble}, \&
  {Avara}}]{ryu2023}
{Ryu}, T., {Krolik}, J., {Piran}, T., {Noble}, S.~C., \& {Avara}, M. 2023,
  \apj, 957, 12, \dodoi{10.3847/1538-4357/acf5de}

\bibitem[{{Saez} {et~al.}(2009){Saez}, {Chartas}, \& {Brandt}}]{Saez2009}
{Saez}, C., {Chartas}, G., \& {Brandt}, W.~N. 2009, \apj, 697, 194,
  \dodoi{10.1088/0004-637X/697/1/194}

\bibitem[{{Saxton} {et~al.}(2012){Saxton}, {Read}, {Esquej}, {Komossa},
  {Dougherty}, {Rodriguez-Pascual}, \& {Barrado}}]{Saxton2012}
{Saxton}, R.~D., {Read}, A.~M., {Esquej}, P., {et~al.} 2012, \aap, 541, A106,
  \dodoi{10.1051/0004-6361/201118367}

\bibitem[{{Schlafly} \& {Finkbeiner}(2011)}]{Schlafly2011}
{Schlafly}, E.~F., \& {Finkbeiner}, D.~P. 2011, \apj, 737, 103,
  \dodoi{10.1088/0004-637X/737/2/103}

\bibitem[{{Shakura} \& {Sunyaev}(1973)}]{shakura73}
{Shakura}, N.~I., \& {Sunyaev}, R.~A. 1973, \aap, 24, 337

\bibitem[{{Shiokawa} {et~al.}(2015){Shiokawa}, {Krolik}, {Cheng}, {Piran}, \&
  {Noble}}]{Shiokawa}
{Shiokawa}, H., {Krolik}, J.~H., {Cheng}, R.~M., {Piran}, T., \& {Noble}, S.~C.
  2015, \apj, 804, 85, \dodoi{10.1088/0004-637X/804/2/85}

\bibitem[{{Steinberg} \& {Stone}(2022)}]{steinberg}
{Steinberg}, E., \& {Stone}, N.~C. 2022, arXiv e-prints, arXiv:2206.10641,
  \dodoi{10.48550/arXiv.2206.10641}

\bibitem[{{Stone} \& {van Velzen}(2016)}]{Stone2016}
{Stone}, N.~C., \& {van Velzen}, S. 2016, \apjl, 825, L14,
  \dodoi{10.3847/2041-8205/825/1/L14}

\bibitem[{{Thomsen} {et~al.}(2022){Thomsen}, {Kwan}, {Dai}, {Wu}, {Roth}, \&
  {Ramirez-Ruiz}}]{Thomsen22}
{Thomsen}, L.~L., {Kwan}, T.~M., {Dai}, L., {et~al.} 2022, \apjl, 937, L28,
  \dodoi{10.3847/2041-8213/ac911f}

\bibitem[{{Tombesi} {et~al.}(2010){Tombesi}, {Cappi}, {Reeves}, {Palumbo},
  {Yaqoob}, {Braito}, \& {Dadina}}]{Tombesi10}
{Tombesi}, F., {Cappi}, M., {Reeves}, J.~N., {et~al.} 2010, \aap, 521, A57,
  \dodoi{10.1051/0004-6361/200913440}

\bibitem[{{Tombesi} {et~al.}(2015){Tombesi}, {Mel{\'e}ndez}, {Veilleux},
  {Reeves}, {Gonz{\'a}lez-Alfonso}, \& {Reynolds}}]{Tombesi15}
{Tombesi}, F., {Mel{\'e}ndez}, M., {Veilleux}, S., {et~al.} 2015, \nat, 519,
  436, \dodoi{10.1038/nature14261}

\bibitem[{{Tonry} {et~al.}(2018){Tonry}, {Denneau}, {Heinze}, {Stalder},
  {Smith}, {Smartt}, {Stubbs}, {Weiland}, \& {Rest}}]{atlas}
{Tonry}, J.~L., {Denneau}, L., {Heinze}, A.~N., {et~al.} 2018, \pasp, 130,
  064505, \dodoi{10.1088/1538-3873/aabadf}

\bibitem[{{Turner} {et~al.}(2007){Turner}, {Miller}, {Reeves}, \&
  {Kraemer}}]{Turner2007}
{Turner}, T.~J., {Miller}, L., {Reeves}, J.~N., \& {Kraemer}, S.~B. 2007, \aap,
  475, 121, \dodoi{10.1051/0004-6361:20077947}

\bibitem[{{van Velzen} {et~al.}(2016{\natexlab{a}}){van Velzen}, {Mendez},
  {Krolik}, \& {Gorjian}}]{vanvelzen2016b}
{van Velzen}, S., {Mendez}, A.~J., {Krolik}, J.~H., \& {Gorjian}, V.
  2016{\natexlab{a}}, \apj, 829, 19, \dodoi{10.3847/0004-637X/829/1/19}

\bibitem[{{van Velzen} {et~al.}(2016{\natexlab{b}}){van Velzen}, {Anderson},
  {Stone}, {Fraser}, {Wevers}, {Metzger}, {Jonker}, {van der Horst}, {Staley},
  {Mendez}, {Miller-Jones}, {Hodgkin}, {Campbell}, \& {Fender}}]{vanvelzen2016}
{van Velzen}, S., {Anderson}, G.~E., {Stone}, N.~C., {et~al.}
  2016{\natexlab{b}}, Science, 351, 62, \dodoi{10.1126/science.aad1182}

\bibitem[{{van Velzen} {et~al.}(2021){van Velzen}, {Gezari}, {Hammerstein},
  {Roth}, {Frederick}, {Ward}, {Hung}, {Cenko}, {Stein}, {Perley}, {Taggart},
  {Foley}, {Sollerman}, {Blagorodnova}, {Andreoni}, {Bellm}, {Brinnel}, {De},
  {Dekany}, {Feeney}, {Fremling}, {Giomi}, {Golkhou}, {Graham}, {Ho},
  {Kasliwal}, {Kilpatrick}, {Kulkarni}, {Kupfer}, {Laher}, {Mahabal}, {Masci},
  {Miller}, {Nordin}, {Riddle}, {Rusholme}, {van Santen}, {Sharma}, {Shupe}, \&
  {Soumagnac}}]{vv21}
{van Velzen}, S., {Gezari}, S., {Hammerstein}, E., {et~al.} 2021, \apj, 908, 4,
  \dodoi{10.3847/1538-4357/abc258}

\bibitem[{{Wevers}(2020)}]{Wevers20}
{Wevers}, T. 2020, \mnras, 497, L1, \dodoi{10.1093/mnrasl/slaa097}

\bibitem[{{Wevers} \& {Ryu}(2023)}]{Wevers23}
{Wevers}, T., \& {Ryu}, T. 2023, arXiv e-prints, arXiv:2310.16879,
  \dodoi{10.48550/arXiv.2310.16879}

\bibitem[{{Wevers} {et~al.}(2017){Wevers}, {van Velzen}, {Jonker}, {Stone},
  {Hung}, {Onori}, {Gezari}, \& {Blagorodnova}}]{Wevers17}
{Wevers}, T., {van Velzen}, S., {Jonker}, P.~G., {et~al.} 2017, \mnras, 471,
  1694, \dodoi{10.1093/mnras/stx1703}

\bibitem[{{Wevers} {et~al.}(2019){Wevers}, {Stone}, {van Velzen}, {Jonker},
  {Hung}, {Auchettl}, {Gezari}, {Onori}, {Mata S{\'a}nchez},
  {Kostrzewa-Rutkowska}, \& {Casares}}]{Wevers19}
{Wevers}, T., {Stone}, N.~C., {van Velzen}, S., {et~al.} 2019, \mnras, 487,
  4136, \dodoi{10.1093/mnras/stz1602}

\bibitem[{{Wevers} {et~al.}(2021){Wevers}, {Pasham}, {van Velzen},
  {Miller-Jones}, {Uttley}, {Gendreau}, {Remillard}, {Arzoumanian},
  {L{\"o}wenstein}, \& {Chiti}}]{Wevers2021}
{Wevers}, T., {Pasham}, D.~R., {van Velzen}, S., {et~al.} 2021, \apj, 912, 151,
  \dodoi{10.3847/1538-4357/abf5e2}

\bibitem[{{Wevers} {et~al.}(2022){Wevers}, {Nicholl}, {Guolo},
  {Charalampopoulos}, {Gromadzki}, {Reynolds}, {Kankare}, {Leloudas},
  {Anderson}, {Arcavi}, {Cannizzaro}, {Chen}, {Ihanec}, {Inserra},
  {Guti{\'e}rrez}, {Jonker}, {Lawrence}, {Magee}, {M{\"u}ller-Bravo}, {Onori},
  {Ridley}, {Schulze}, {Short}, {Hiramatsu}, {Newsome}, {Terwel}, {Yang}, \&
  {Young}}]{Wevers2022}
{Wevers}, T., {Nicholl}, M., {Guolo}, M., {et~al.} 2022, \aap, 666, A6,
  \dodoi{10.1051/0004-6361/202142616}

\bibitem[{{Wu} {et~al.}(2018){Wu}, {Coughlin}, \& {Nixon}}]{wu18}
{Wu}, S., {Coughlin}, E.~R., \& {Nixon}, C. 2018, \mnras, 478, 3016,
  \dodoi{10.1093/mnras/sty971}

\bibitem[{{Yao} {et~al.}(2022){Yao}, {Lu}, {Guolo}, {Pasham}, {Gezari},
  {Gilfanov}, {Gendreau}, {Harrison}, {Cenko}, {Kulkarni}, {Miller}, {Walton},
  {Garc{\'\i}a}, {van Velzen}, {Alexander}, {Miller-Jones}, {Nicholl},
  {Hammerstein}, {Medvedev}, {Stern}, {Ravi}, {Sunyaev}, {Bloom}, {Graham},
  {Kool}, {Mahabal}, {Masci}, {Purdum}, {Rusholme}, {Sharma}, {Smith}, \&
  {Sollerman}}]{Yao2022}
{Yao}, Y., {Lu}, W., {Guolo}, M., {et~al.} 2022, \apj, 937, 8,
  \dodoi{10.3847/1538-4357/ac898a}

\end{thebibliography}
\bibliographystyle{aasjournal}

\end{document}